\documentclass[aps,11pt,eqsecnum, preprint,nofootinbib,superscriptaddress]{revtex4}
\usepackage{amssymb,amsmath,amsthm,graphicx,amscd}
\usepackage{enumerate,xcolor,changepage,ulem,bm}
\usepackage[hidelinks]{hyperref}

\DeclareMathSymbol{\shortminus}{\mathbin}{AMSa}{"39}

\begin{document}


\title{Intrinsic Entropy of Squeezed Quantum Fields  and Nonequilibrium \\ Quantum Dynamics of Cosmological Perturbations}

\author{Jen-Tsung Hsiang}
\email{cosmology@gmail.com}
\affiliation{Center for High Energy and High Field Physics, National Central University, Chungli 32001, Taiwan, ROC}
\author{Bei-Lok Hu}
\email{blhu@umd.edu}
\affiliation{Maryland Center for Fundamental Physics and Joint Quantum Institute,  University of Maryland, College Park, Maryland 20742, USA}
\date{\today}

\begin{abstract}
Density contrasts in the universe are governed by scalar cosmological perturbations which, when expressed in terms of gauge-invariant variables, contain a classical component from scalar metric perturbations  and a quantum component from inflaton field fluctuations. It has long been known that the effect  of cosmological expansion on a quantum field amounts to squeezing.  Thus the entropy of cosmological perturbations can be studied by treating them in the framework of squeezed quantum systems. Entropy of a free quantum field is a seemingly simple yet subtle issue.  In this paper, as different from previous treatments,  we tackle this issue with a fully developed nonequilibrium quantum field theory formalism  for such systems. We  compute the covariance matrix elements of the parametric quantum field and solve for the evolution of the density matrix elements and the Wigner functions, and, from them, derive the von Neumann entropy. We then show explicitly why the entropy for the squeezed yet closed system is zero, but is proportional to the particle number produced upon coarse-graining out the correlation between the particle pairs. We also construct the bridge between our quantum field-theoretic results and those using probability distribution of classical stochastic fields by earlier authors. From this we can see the clear advantages of the quantum field-theoretical approach over the stochastic classical field treatment since the latter misses out in some important quantum properties, such as  entanglement and  coherence, of the quantum field.  \\

-- {\it Invited paper in Entropy: Special Issue on “Entropy Measures and Applications in Astrophysics”}
\end{abstract}


\maketitle

\hypersetup{linktoc=all}
\setcounter{tocdepth}{2}

\clearpage
\baselineskip=18pt
\numberwithin{equation}{section}
\allowdisplaybreaks

\section{Introduction}

Entropy of quantum cosmological perturbations is an important topic which has by now clocked almost three decades of investigations \cite{BMP,Prok93}.
In terms of its theoretically foundation, it is built upon the bigger issue of  {\bf I) the entropy of quantum fields}, where investigation started in the mid-80s~\cite{HuPav,HuKan,HabKan,CalHu88} continuing on to the 90s~\cite{GasGio,InfSqV,CalHu95},  00s~\cite{CalHu00,CalHu03,KiePS00,AMM05} and rekindled in  recent years~\cite{KPS12,CamPar,Boyan,Burgess,Brahma}.
There are two aspects in this theme,  1) the {\it quantum field theory} component depicting particle creation from the vacuum and quantum cosmological perturbations;  2)  the  {\it nonequilibrium statistical mechanics} aspect describing the evolutionary dynamics of a quantum many-body system, the  fluctuations in, and the dissipation of, a quantum field. These two components when combined make up quantum field theory of nonequilibrium systems~\cite{NEqFT}, the  two major paradigms being the Boltzmann correlation hierarchy and the Langevin open systems~\cite{CalHu08}.  These two components or aspects enter in many theoretical issues of fundamental interest  such as quantum (de)coherence, quantum correlations and quantum entanglement. Concepts in open quantum systems~\cite{qos} have been introduced and advanced techniques in quantum field theory~\cite{NEqFT,CalHu88} utilized for these purposes~\cite{CalHu08}.  In this backdrop we wish to investigate the entropy of quantum fields and cosmological perturbations in a dynamical setting, such as in the early universe.  Saving the discussions of the  two theoretical factors for  the next section we focus here on the cosmological aspect.  {\bf  II)  The cosmological factor} also has two components. They are: 1) {\it Gravitational perturbation} theory which describes how weak classical perturbations of scalar (density contrast) vector (vorticity) and tensor (gravitational waves) types evolve in a dynamical spacetime;  and 2) {\it Quantum matter field}  processes such as particle creation from vacuum fluctuations amplified by the expansion of the universe and their consequences. 

\subsection{The Cosmological Aspect: Gravitational Perturbations and Quantum Fluctuations}

\subsubsection{Classical gravitational perturbations: scalar, vector and tensor components} 

Gravitational perturbation theory is a well established subject  in cosmological structure formation since the 1946 seminal paper of Lifshitz based on the amplification of density contrasts related to the (scalar sector of the) metric perturbations \cite{Lif46,LK63,Haw66,PeebYu,Bardeen}.  What seeded the structures in the classical gravitational perturbation theory was assumed to be from white noises.  After the advent of inflationary universe \cite{Guth,LindeBook,MukhanovBook} where the vacuum energy density of a quantum scalar field, the inflaton, is believed to have driven the cosmos to (near-) exponential expansion for a certain duration in the early universe, one needs to take into account how the quantum scalar field and its fluctuations are coupled to the (scalar sector of the) metric perturbations controlling the density constrasts.  If we regard Lifshitz's 1946 paper as opening the first (classical) stage of cosmological structure formation investigtion based on  gravitational perturbation theory, this second (quantum) stage involving the inflaton's fluctuations began in 1982 \cite{GuthPi82,Haw82,Staro82,BST,FMB,LytLid09}.   

With the attention now focused on everything we see today in the universe as originating from quantum field mediated-{\it gravitational perturbations}, it is worth mentioning  another important  processes involving  {\it fluctuations of quantum fields} in the early universe,  specifically, cosmological particle creation from the vacuum.    The former class of activities placed in the inflationary universe context could take place as early as the GUT time ($10^{-35}$ sec) while the latter happened even earlier, predominantly at the Planck time ($10^{-43}$ sec).

\subsubsection{Quantum field processes involving vacuum fluctuations} 

Quantum field processes in curved spacetime such as Hawking radiation from  black holes or cosmological particle creation in the very early universe are described in great details in many monographs, e.g., \cite{BirDav,Fulling,Wald,ParTom,HuVer20}. 
The amplitude  of a classical wave mode can be parametrically amplified by a time-dependent drive \cite{Zel70}.  Same way for vacuum fluctuations in a quantum field, resulting in the creation of particle pairs:  the expansion of the universe acting like a drive, parametrically amplifying the quantum noise, giving rise to spontaneous particle production \cite{Par69,Zel70}.  If there were particles already present in an initial state they will get amplified  with the same amplification factor in stimulated production.  This is like the spontaneous and stimulated emission of atoms in quantum optics. In fact,  quantum field processes in  a time-dependent background, be it in a cosmological spacetime  or in an external laboratory field \cite{QFText},  can be captured  by  the `squeezing' of quantum states \cite{Walls,LouKni,ManWol}. The vacuum is  `squeezed' in the evolutionary history while particles are produced  \cite{GriSid}. 
A summary description of cosmological particle creation in terms of squeezing can be found in e.g., \cite{HM94,HKM94,InfSqV}. 

 We mention quantum cosmological perturbations and cosmological particle creation together because they are subjected to the same mechanism which amplifies these perturbations or fluctuations,  namely,  as in parametric oscillators,  where the frequencies of the normal modes are time-dependent.  While parametrically amplified quantum fluctuations engender particle creation, parametrically amplified gravitational perturbations engender galaxies and structures,  either classically with seeds of  white noise or  by  the inflaton field's quantum fluctuations.

\subsubsection{Distinguish classical perturbations from quantum fluctuations}

It is of theoretical significance to make the distinction between classical linear gravitational perturbations,  which are believed to be the progenitors of galaxies and structures we see today,    and vacuum fluctuations of a quantum field, which engender spontaneous creation of particle pairs, a subject fundamental in quantum field theory in curved spacetime. When we talk about the cosmological density contrasts, the isocurvature perturbations, the vorticity and the primordial gravitational waves, we are referring to quantities derived from the scalar, vector and tensor {\it perturbations} of the background spacetime.
Density contrasts are derived from the scalar sector of the metric perturbations, a subject well explored in classical general relativity, culminating in Bardeen's gauge invariant quantities \cite{Bardeen,FMB}.  In inflationary cosmology they are coupled to a quantum scalar field, the inflaton.  To get one compact equation  of motion  for the density contrast,  mixed metric perturbations + scalar field variables  are used, such as the gauge invariant Mukhanov-Sasaki variable.  This is all standard and fine.  But when one moves to the quantum theory of  cosmological perturbations in inflationary universe, one has to be careful what this means, especially when dealing with decoherence and quantum to classical transition issues.  The scalar (inflaton) field $\hat \Phi ({\bf x}, t)$ is intrinsically quantum in nature and comes with its quantum fluctuations.  Often a background field expansion $\hat \Phi ({\bf x}, t) = \bar \phi ({\bf x}, t)+ \hat \varphi ({\bf x}, t)$ is performed,  assuming the background field $\bar \phi ({\bf x}, t)$ is classical and  the quantum character shows through its fluctuations $\hat \varphi ({\bf x}, t)$.   However,  the gravitational perturbation component in this mixed variable is of classical origin. The scalar sector of the metric perturbations related to the Newtonian potential is a constraint, not a dynamical degree of freedom (the tensor modes, the gravitational waves,  are).  Its nature is determined by (or `slaved'  to \cite{AnaHuNSE,AnaHu2Cat} ) the matter source. In general relativity when the matter source is classical, this scalar sector of the metric perturbation is classical.  In inflationary cosmology,  what determines the density contrast comes from both the classical scalar metric pertubations and the quantum inflaton field. When one says, `quantize the mixed variable', one should bear in mind that the intrinsic nature of metric perturbations remains classical.  In an extreme case, one may even conjure up situations where the quantum  fluctuations of the scalar field are made to vanish, such as ``choosing a  (co-moving) gauge which for scalar perturbations makes the velocity perturbation vanish.  For single field inflation,  this means that the time coordinate is defined so that at any given time the scalar field equals its unperturbed value" (one is riding up and down with the scalar field's fluctuations),  ``with all perturbations relegated to components of the metric.'' (\cite{Weinberg}, Sec. 5.3D)  This does not mean that gravity has become quantum, only that the scalar perturbations now acquire a quantum nature by virtue of the presence of the inflaton field.  Put it in another way, when there is no inflaton, one returns to purely classical general relativity.  The Newtonian force is slaved to the source which is classical. There is no way for the gravitational perturbations to become quantum.  

Now about the tensor sector.    Gravity's dynamical (or propagating) degrees of freedom reside only in the tensor sector, i.e., the gravitational waves.  Primordial gravitational waves are described by the tensor component of gravitational metric perturbations. They have also been studied at the classical level since 1946.  One can consider quantizing the  linearized tensor perturbations, whence they become the  primordial gravitons\footnote{Note gravitational waves are weak metric perturbations, like sound wave (barring the differences between transverse and longitudinal waves). Gravitons are quantized linear perturbations of spacetime,  like phonons,  in the nature of collective excitations.  This is a far cry from quantum gravity, defined as theories for the microscopic structures of spacetime at the Planck scale \cite{E/QG}}.   These gravitons still obey deterministic equations of motion.  They are {\it not} stochastic intrinsically.  Only when one considers a large number of primordial  gravitons  and use statistical means to describe their distributions would they become `stochastic'  \footnote{Probing further into their connection in cosmological perturbation theory Roura and Verdaguer~\cite{RouVer} show that in the theoretical framework of stochastic gravity~\cite{HuVer20}, at the Gaussian level,  the stochastic variables   give an equivalent description as the quantized linear perturbations~\cite{FMB}}. 


 
\subsection{Related issues:  Decoherence and Entanglement}

Before we delve into a full disucssion of entropy of cosmological perturbations and fluctuations we want to mention two issues in quantum information related to entropy, namely,  decoherence and entanglement. This would be brief becasue each topic would merit a separate paper or two to describe.  


1) {\it Decoherence of cosmological perturbations}. Again, many questions are asked and different approaches have been suggested. To give two early examples, in Guth and Pi \cite{GuthPi85}, by using an inverted harmonic oscillator model as example and invoking the uncertainty relation they showed that the inflaton will at late times behave classically.  Note that this is for a closed system thus strictly speaking it is at most dephasing,  rather than decoherence.  In Starobinsky's stochastic inflation model \cite{StoInf}, two key points:  a) the long wavelength perturbations are assumed to behave classically and b) the short wavelength perturbations are treated  as white noise.  Under what conditions could a) be realized and b) be implemented?  How could it be that a free field partitioned into two sectors in an expanding universe that the short wavelength sector would behave like white noise? One should also clarify the  conditions this noise can effectively decohere  the long wavelength sector by its backreaction.  
These questions were asked and challenged by, e.g.,  \cite{CalHu95,VilWin}.  This area of research begun in earnest in the early 90s \cite{BLM,HPZdec,CalHu95,LomMaz,Matacz97,Kiefer99} is still being pursued vibrantly today. For a sample of recent work,  see, e.g., \cite{Burgess,Brahma},  where other earlier references can be found.  We shall discuss this topic with a detailed background summary in a companion paper under preparation \cite{HHCosDec}.

2) {\it Entanglement and entropy}.   The entanglement between particle pairs, one with momentum ${\bf k}$, the other with $- {\bf k}$, such as studied in \cite{CamPar,LCH10} is relevant to our present consideration of entropy associated with particle creation.  Another related topic is  {\it entanglement entropy}.  The seminal papers  \cite{BomSor,Sred}  explore  the entropy of free quantum fields in Minkowski space with a \textit{partition},  providing a more general and basic statistical mechanical way to understand black hole entropy.  It was the precursor of the by-now familiar topic of entanglement entropy \cite{EntEnt} hotly pursued in the last two decades.  In cosmology when the spacetime is dynamical the natural `partition' which separates modes into two sectors is the Hubble horizon defined by the inverse of the expansion rate.  Entanglement between two Unruh-DeWitt detectors has been studied concerning the effects of spacetime curvature~\cite{MarMenCosEnt,Terno} and topology \cite{LCH16} and under other conditions in the emergent field of relativistic quantum information \cite{RQI}, but that is not the main concern in our present study.  We now turn to the two aspects mentioned in the beginning, that of quantum field theory and nonequilibrium dynamics.


\section{Entropy of Quantum Fields}

Defining and understanding entropy for quantum fields is a task of fundamental significance in both quantum field theory and nonequilibrium statistical mechanics. For free fields in a dynamical setting (driven by some external source or in a dynamical spacetime as in cosmology) with no  spatial boundary or  event horizon present (thus not involving entanglement entropy) the authors of \cite{HuPav} provided a first answer to the following question: Is there entropy production in particle creation from the vacuum? Adhering to quantum field theory,  because the vacuum is a pure state and particle creation is a unitary process, the answer should be no.  However,  even in textbooks one sees that an entropy, say, of photons,  can be assigned proportional to the number of particles present. These two seemingly contradictory answers each seem to stand on its firm ground. How does one reason out their differences? This paradox was what lured one of us into exploring this issue and came out with some interesting discoveries. The first observation is:  the first answer  seems to focus on the initial state while the second answer on the final outcome. For the second answer to make sense some information must have been lost. Two possibilities come to one's  mind. Either a)  some essential information is ignored, implicitly when one defines the field entropy in terms of particle numbers,  or b) some kind of coarse-graining measure is introduced explicitly which curtails some information of the system. 

For a), when one argues that the entropy of a free quantum field is  proportional to the particles created,  one often  implicitly adopts a  Fock space representation to capture the particle number,  while ignoring completely the information about the correlation and the coherence in the particle pairs. This was pointed out in \cite{HuPav}.  Note that the entropy of a closed system is defined by taking the full trace, which is independent of the basis or representation, and there is no priori definition or requirement that the entropy be a function of the particle number.  This point is illustrated explicitly in \cite{KME97} where an equation governing the quantum coherence is presented alongside with that governing the particle numbers, from which the entropy is calculated. For b), in a realistic setting, after particles are created from the vacuum, they interact, and information about their interaction could be lost, e.g., if one focuses on the scattering cross section but ignores some of their correlations. This second point is illustrated in an interacting harmonic oscillator model of field theory in \cite{HuKan} and later with greater depth  in \cite{CalHu03} which proves an $H$-theorem in such coarse-grained systems. 



\subsection{Intrinsic Entropy of Free Quantum Fields} 

Entropy of quantum fields is a fundamental, seemingly simple yet somewhat tricky issue. Early conceptual inquiries \cite{HuPav,HuKan} made explicit the specific underlying conditions which allow for the entropy of free quantum fields to be related to particle creation numbers (for boson fields). Namely, by adopting a Fock space representation and making statements only in terms of number operators, one implicitly ignores all quantum phase information which determines the coherence. (See \cite{KME97} for the interplay of both quantities). From this it is easy to understand why the entropy associated with particle production is proportional to the degree of squeezing, in cosmology by the expansion of the universe \cite{GasGio,Prok93,KMH97}, or more down to earth, the moving mirror  in dynamical Casimir effect \cite{DCE}. 



\subsection{Entropy of Interacting Quantum Fields}

We have alluded to this class of theories above, the simplest exemplified by the anharmonic oscillator models in quantum mechanics. A more systematic analysis is based on the Boltzmann paradigm  applied to the BBGKY hierarchy, namely,  the `slaving' of the higher order correlation functions to the lower ones, with Boltzmann equation being the lowest order,  describing the dynamics of the one point distribution function with source from the two particle collision integral.  The assumption of a causal factorizable initial condition on the two-particle correlation function  gives rise to a dissipative dynamics in the one particle distribution function.    For interacting quantum field,  it is the concept of {\it correlation entropy} defined in the Schwinger-Dyson hierarchy for a certain $n$th order  correlation functions. See \cite{CalHu93,cddn,CalHu00,CalHu03}. Notice the fundamental difference between the `truncation' and the `slaving' procedures, the former giving rise to a unitary equation such as for the mean field in the Vlasov equation, the latter a (nonunitary) dissipative Boltzmann equation.    

\subsection{Entropy Measures} 

As we stated in the beginning and is well known, entropy arises from a loss of information, either by choice or by necessity. Thus, any mention of  entropy measure  must be accompanied by  the specification of what information is being dropped, lost,  or operationally inaccessible.  That in turn depends on one's coarse-graining choices or one's level of ignorance.  We give  as example two commonly used coarse-graining measures. 

\subsubsection{Coarse-Graining by Discrepancies of Scales }

If the relevant physical variables in a system have clear discrepancies in scales, be it slow vs fast, low vs high frequency, light vs heavy mass,  one can decide based on physical conditions which scale is of the most physical interest and proceed to coarse-grain the other variables, one level at a time  in an orderly nested way.   Famous examples are the slow variables of van Hove, the adiabatic invariants for a parametric process,  the Born-Oppenheimer approximation wherein the  heavy mass (e.g., nucleus) motion  is considered as effectively static with respect to the light mass (e.g., electron), etc. Formally,  these situations are best described by  the projection operator formalism of Zwanzig-Mori-Nakajima \cite{Zwanzig}.  In quantum systems, one measure often adopted is the coarse-graining of the phase information since the phase often varies rapidly. This is behind the random phase approximation.  The rotating wave approximation (RWA) is one of the two preambles of quantum statistical mechanics, where the density matrix of a quantum system is assumed to be diagonal. By eliminating all phase information one can consider only probabilities, which is why quantum statistical mechanics taught from textbooks misses all the quantum coherence and entanglement issues. 

\subsubsection{Coarse-graining by Partitioning}

If there is a clear cut partition in the system one is interested in, one can use it to separate the variables of special interest from other variables  perhaps of lesser interest.  In the stochastic inflation example, the horizon is the divider between the long wavelength sector and the short wavelenght sector. The physical issues of concern are  accordingly different: for the former, decoherence --  how could they be viewed as classical? For the latter, noise --  do they really constitute  white noise?   For interacting quantum fields, if there is only one field and one has chosen the appropriate partition one can invoke the  coarse-grained effective action (CGEA) \cite{cgea,JH1,CHM00}   to describe the backreaction effects of the coarse-grained sector viewed as the environment to the system. The  `in-in', Schwinger-Keldysh, or `closed-time-path' version of CGEA is equivalent to the Feynman-Vernon influence functional formalism. This approach was used to describe cosmological decoherence in \cite{HPZdec,CalHu95,LomMaz}.  If there are more than one field, it is even easier,  because the system field and the environment field are distinct and their coupling does not change in time. One can decide which field should be considered as one's system and coarse-grain the others \cite{ZhangPhD,Banff,Boyan}. One does not need to keep track of the time-varying partition in an expanding universe.
  
 In cosmology,  when the spacetime is dynamical,  the natural `partition' is between the modes of wavelengths greater or less than the horizon,  defined by the inverse of the Hubble expansion rate --  between the super and sub horizon sectors.  This is referred to as entanglement entropy in cosmology.   Recent work include   \cite{ProRig,HolMcD17,Nelson,Brahma,RaiBoy}, where earlier references can be found.

\subsection{Approaches taken here in relation to earlier work}

From the pioneering work in the 80s and 90s mentioned above it is encouraging to see that the conceptual framework of open quantum systems and the  technical tools of nonequilibrium quantum field theory are more widely adopted in current theoretical cosmological research. For recent work see, e.g., \cite{Fukuma,Boyan,Burgess,HolMcD17,Nelson,Brahma} and references therein.  We now focus on the specific main points on the entropy of cosmological perturbations, analyzed here from a squeezed open quantum system perspective using techniques  from nonequilibrium quantum field theory and relate to earlier work.

1) The {\bf  theoretical framework} useful for our purpose is known as `squeezed open quantum systems' \cite{HM94,HKM94,KMH97} with cosmology being an important exemplary category \cite{InfSqV}. For the {entropy measure} we shall follow 

A)  the vein of  \cite{HuPav,KME97}  where, using a Fock basis  measuring the number of particles created, or the cosmological perturbations,  we show explicitly when the quantum correlations between the particle pairs are kept,  there is no entropy generation, whereas when the correlation is ignored,  we obtain the well known expression for the entropy proportional to the number of particles.  Related to \cite{KME97} is the work of \cite{AMM05} where the entropy is calculated from the  von Neumann entropy constructed from the reduced density matrix defined with respect to a  second adiabatic order number state\footnote{Of interest is their conclusion that there is no decoherence as claimed in \cite{PolSta96} (despite their `cute'  way of explaining why there {\it is} decoherence).}. 

B) Our demonstration  that there is entropy generated   by coarse-graining the $-\bm{k}$ component of the particle pair corroborates the results obtained by \cite{LCH10} via a different pathway,  by showing the destruction of the entanglement between the particle pair.  


3)  The {\bf formal structure}.  Many authors,  including those of the major work \cite{BMP} (BMP) and \cite{CamPar} (CP) on the entropy of cosmological perturbations,  correctly start with the formal structure of an interacting quantum field theory.   Treating  interacting fields in  the framework of the classical BBGKY or the quantum Schwinger-Dyson hierarchy -- the Boltzmann paradigm -- is a serious challenge.  It was taken up in earnest in the work of  \cite{CalHu93,CalHu95,CalHu00,CalHu03},  developing the concepts of correlation histories, correlation dynamics and correlation noise.  Their master effective action and the way how the  higher order correlators should be `slaved' to a lower order correlator provide  the basis for formulating effective field theories  for open quantum systems \cite{CalHu97}, an important theme which saw fruitful applications to cosmological issues in \cite{Boyan}.   In the same vein,  entropy from non-Gaussian states is pursued  by  \cite{KPS12}. 
However, BMP  and  CP  quickly assume   a Gaussian truncation of the BBGKY or Schwinger-Dyson hierarchy, thus restricting their attention only to the lowest two-point functions.   This is enough  to treat linear perturbations  and effectively places  their theories  on the same footing as free fields, where considerations in \cite{HuPav,LCH10} would suffice.    CP refer to the reduced density matrices defined by truncating the hierarchy at the first level as Gaussian and homogeneous density matrices (GHDM). Coarse-graining  the phase information is also used in two major work:  In  \cite{BMP}   their coarse-grained  entropy is obtained by averaging the two point correlation function over the phases. In \cite{CamPar}  the growth of entropy associated with the coarse-graining is described by the set of Gaussian two-mode reduced density matrices $\rho^{red}_{({\bm k}, -{\bm k})}$,  each  characterizes the loss of entanglement between the two modes in the presence of interactions.    

This is  also the starting point of the way  BMP define entropy for gravitational field.  However,  instead of treating a quantized   field,  they consider  a classical stochastic field.  Thus their considerations are predicated upon the quantum perturbations having effectively decohered: 
``Provided there is  decoherence and $\sinh r_{\bm{k}} \gg 1$ (where $r_{\bm{k}}$ is the squeeze parameter of the $\bm{k}$th mode),  then we can take the classical limit where the classical correlation functions can be identified with quantum expectation values, i.e., $\langle\phi(x)\phi(y)\rangle = \langle 0\vert\hat{\phi}(x) \hat{\phi}(y)\vert0\rangle$''. Note the angular brackets on the lefthand side denotes a statistical average.  The decoherence procedure is clearly stated in \cite{Prok93} as 
dropping the off-diagonal components of the density matrix.
At the time these papers were written decoherence as an important theoretical issue was not that well understood. See more recent papers such as  \cite{KPS12,Brahma} and \cite{HHCosDec} where the developments in the last 30 years can be traced back.   We have devoted a subsection at the end of this paper to clarify the basic differences between a quantum field theory treatment  and a classical stochastic field theory treatment.

\subsection{Our Findings and Organization}

Our narrative takes on four steps, showing:  1) that the cosmological expansion results in the squeezing of the quantum field,  2) the nonequilibrium  dynamics of a squeezed quantum field as a closed system, seen through the evolution of the density matrix operator  in terms of the covariance matrix elements,  3) that the von Neumann entropy is zero in this closed system, as expected, and proportional to the particle numbers when the quantum correlation between the particle pair is coarse-grained away. We derive 4) the Wigner function of the quantum Gaussian field driven by a parametric process, which gives a clear picture of how the coarse-grained entropy may emerge, and whose phase-space description leads to the quantum thermodynamics of this system. And, with this we also  discuss 5) the differences between our quantum field-theoretical methodology and approaches in the literature using  the probability distribution of   classical stochastic fields, the latter formulation cannot treat quantum coherence and entanglement issues.  While facts  in 1) and 3) are largely known before,  we try to capture a global perspective and point out the subtleties in the meaning of entropy and how it is defined in closed quantum systems. Points  2), 4), and 5) form the technical core and the results we obtained are new.

This paper is organized as follows:  in Section \ref{S:euibd}, to be self-contained, we give a quick overview of the essential ingredients of  first-order classical cosmological perturbation theory, and link the cosmological perturbations to the classical parametric field via the Mukhanov-Sasaki variables. In Sec.~\ref{S:ebhf} we formally compute the covariance matrix elements of the parametric field as  the basic building blocks of a Gaussian system. Then in Sec.~\ref{S:ebsfg} we decompose the field into a collection  of parametric oscillators to take advantage of the oft-used  nonequilibrium formalism to construct the density matrix operator and the density matrix elements.  Doing so equips us to tackle the full dynamical evolution of the physical observables and derive the nonequilibrium thermodynamics of the system under study. We present the time evolution of the density matrix and its elements in Sec.~\ref{S:ebgf} where we combine the language of the Bogoliubov transformation and the aforementioned nonequilibrium formalism for a succinct description of the full dynamics. In Sec.~\ref{S:roithnd}, we show  entropy  production associated with the particle pairs in the parametric development of the cosmological perturbations. In Sec.~\ref{S:wigner}  we construct the Wigner function of the quantum field and show how it evolves in time.  As a bonus we also show the relation between our quantum field theoretic results and those using probability distribution of classical stochastic fields by earlier authors. We point out the clear advantage of the former as the latter misses out in some core quantum properties, such as quantum entanglement and quantum coherence of the field.

\section{cosmological perturbation theory in a nutshell}\label{S:euibd}

In this short section we give a brief summary of cosmological perturbation theory~\cite{cosPer} to make our presentation self-contained and to define the notations,  which have evolved to become quite standardized over the years.  Readers familiar with this topic can skip to the next section. 

\subsection{Gauge transformations}

The linear perturbations $\delta g_{\mu\nu}$ off a background spacetime with metric $\bar{g}_{\mu\nu}$.  In the case of a 
FLRW universe background with scale factor $a$ we can write the metric as
\begin{equation}
	g_{\mu\nu}=\bar{g}_{\mu\nu}+\delta g_{\mu\nu}=a^{2}\bigl(\gamma_{\mu\nu}+h_{\mu\nu}\bigr)\,,
\label{MetricPert}
\end{equation}
where $h_{\mu\nu}$ and its derivatives are assumed to be small compared to the background metric $\gamma_{\mu\nu}$, which is used for the raising and lowering of the indices for the metric perturbations. Namely, 
\begin{align}
	h^{\mu}{}_{\nu}&=\gamma^{\mu\alpha}h_{\alpha\nu}\,,&h^{\mu\nu}&=\gamma^{\mu\alpha}\gamma^{\nu\beta}h_{\alpha\beta}\,,
\end{align}
such that the inverse metric of the perturbed universe is
\begin{equation}
	g^{\mu\nu}=a^{-2}\bigl(\gamma^{\mu\nu}-h^{\mu\nu}\bigr)\,.
\end{equation}
 The metric perturbations $h_{\mu\nu}$ can be decomposed into the scalar, vector and tensor components, but  the scalar component of the perturbations is essential for the consideration  of cosmological density contrasts.  The line element can therefore be parametrized as
\begin{equation}
	ds^{2}=a^{2}(\eta)\Bigl\{-\bigl(1+2\mathcal{A}\bigr)\,d\eta^{2}-2\mathcal{B}_{,\,i}\,d\eta dx^{i}+\bigl[\bigl(1-2\psi\bigr)\,\delta_{ij}+2\partial_{i}\partial_{j}\mathcal{E}\bigr]\,dx^{i}dx^{j}\Bigr\}\,,
\end{equation}
by four scalars $\mathcal{A}$, $\mathcal{B}$, $\psi$ and $\mathcal{E}$. Here $\eta$ is the conformal time coordinate and $x^{i}$ are the spatial coordinates. 

Now consider the choice of gauge and the construction of gauge-invariant variables, 
the infinitesimal coordinate transformation between two coordinate systems, say $\{x^{\alpha}\}$ and $\{\tilde{x}^{\alpha}\}$, in the perturbed spacetime is given by
\begin{equation}\label{E:bvgerd}
	\tilde{x}^{\alpha}=x^{\alpha}+\xi^{\alpha}\,,
\end{equation}
where $\xi^{\alpha}$ and their derivatives are first-order small in $h$.  It then can be shown that the four scalars in the line element will transform as
\begin{align}
	\tilde{\mathcal{A}}&=\mathcal{A}-\xi^{0}{}'-\frac{a'}{a}\,\xi^{0}\,,&\tilde{\psi}&=\psi+\frac{a'}{a}\,\xi^{0}\,,&\tilde{\mathcal{B}}-\tilde{\mathcal{E}}'&=\mathcal{B}-\mathcal{E}'+\xi^{0}\,,
\end{align}
where the prime denotes the time derivative and $\xi$ is the scalar component of the spatial vector $\xi^{i}$. Among them, we can form two gauge-invariant Bardeen potentials
\begin{align}
	\Phi&=\mathcal{A}+\frac{a'}{a}\bigl(\mathcal{B}-\mathcal{E}'\bigr)+\bigl(\mathcal{B}-\mathcal{E}'\bigr)'\,,&\Psi&=\psi-\frac{a'}{a}\bigl(\mathcal{B}-\mathcal{E}'\bigr)\,.
\end{align}
which are invariant under the coordinate transformation \eqref{E:bvgerd}, and $\tilde{\Phi}={\Phi}$ and $\tilde{\Psi}={\Psi}$. If we choose $\xi=-\mathcal{E}$ and $\xi^{0}=-(\mathcal{B}-\mathcal{E}')$, then we have $\tilde{\mathcal{B}}=0$ and $\tilde{\mathcal{E}}=0$. This is the Newtonian gauge, in which $\tilde{\mathcal{A}}=\tilde{\Phi}$, and $\tilde{\psi}=\tilde{\Psi}$ become gauge-invariant. In terms of the gauge-invariant quantities the line element    reduces to 
\begin{equation}
	ds^{2}=a^{2}\Bigl\{-\bigl(1+2\Phi\bigr)\,d\eta^{2}+\bigl(1-2\Psi\bigr)\delta_{ij}\,dx^{i}dx^{j}\Bigr\}\,.
\end{equation}
 The Bardeen potentials will be equal if the stress tensor of matter has no anisotropic perturbations.

\subsection{Scalar field perturbations}

Next we turn to the perturbation of a real scalar field
\begin{equation}\label{E:gjesresd}
	S_{\phi}=-\int\!d^{4}x\sqrt{-{g}}\;\,\Bigl\{\frac{1}{2}\,{g}^{\mu\nu}\partial_{\mu}\phi\partial_{\nu}\phi+V(\phi)\Bigr\}\,,
\end{equation}
where $V$ is its potential. We will assume that the unperturbed spacetime is homogeneous and isotropic, and that the scalar field in this spacetime is a function of time only. The metric perturbations will induce perturbations of the field. The scalar field perturbations $\delta\phi$  obey  the wave equation
\begin{equation}\label{E:gjvsjdf}
	\delta\phi''-\partial^{2}\delta\phi+2\mathcal{H}\delta\phi'+a^{2}\,\frac{\partial^{2} V(\varphi)}{\partial\varphi^{2}}\,\delta\phi=\bigl(\mathcal{A}'+3\mathcal{D}'+\partial^{2}\mathcal{B}\bigr)\varphi'-2a^{2}\mathcal{A}\,\frac{\partial V(\varphi)}{\partial\varphi}\,,
\end{equation}
where $\mathcal{H}=a'/a$ is the Hubble expansion rate  in conformal time. We see that the scalar field perturbations couple only to scalar metric perturbations.

After  some  algebra, one can derive the Einstein equations for the scalar perturbations given by
\begin{align}
	\partial^{2}\Phi-3\mathcal{H}\Phi'-3\mathcal{H}^{2}\Phi&=4\pi a^{2}\Bigl[\frac{1}{a^{2}}\Bigl(\varphi'\,\delta\phi'-\varphi'^{2}\Phi\Bigr)+\frac{\partial V(\varphi)}{\partial\varphi}\,\delta\phi\Bigr]\,,\label{E:ertsrgf1}\\
	\Phi'+\mathcal{H}\Phi&=4\pi\varphi'\delta\phi\,,\label{E:ertsrgf2}\\
	\Phi''+3\mathcal{H}\Phi'+\bigl(2\mathcal{H}'+\mathcal{H}^{2}\bigr)\Phi&=4\pi a^{2}\Bigl[\frac{1}{a^{2}}\Bigl(\varphi'\,\delta\phi'-\varphi'^{2}\Phi\Bigr)-\frac{\partial V(\varphi)}{\partial\varphi}\,\delta\phi\Bigr]\,,
\end{align}
with
\begin{equation}
	\varphi''+2\mathcal{H}\varphi'+a^{2}\frac{\partial V(\varphi)}{\partial\varphi}=0\,.
\end{equation}
Recall that $\Phi$ is the Bardeen potential.  Taking a suitable superposition of the metric perturbation and the scalar field perturbation, Mukhanov and Sasaki define  the gauge-invariant variable $u$  in the conformal Newtonian gauge as
\begin{equation}
	u=\frac{a^{2}}{a'}\,\varphi'\Bigl(\Phi+\frac{a'}{a\varphi'}\delta\phi\Bigr)\, 
\end{equation} 
and  rewrite the Einstein equation   in the conformal Newtonian gauge into the Mukhanov-Sasaki equation
\begin{align}
	u''-\partial^{2}u-\frac{z''}{z}\,u&=0\,,&&\text{with}&z&=\frac{a^{2}}{a'}\,\varphi'\,.\label{E:oqasdfs}
\end{align} 

Eq.~\eqref{E:oqasdfs} is in the form of an equation of motion for a parametric field $u(\bm{x},\eta)$ in flat space. We shall focus on this equation and discuss the dynamics of a generic parametrically-driven quantum scalar field in the next sections.

\section{Dynamics of a parametrically-driven quantum field}\label{S:ebhf}

We consider a real, massive scalar parametric field $\phi(\bm{x},t)$ in a flat spacetime. We assume that the field has a time-dependent  mass $m(t)$, which changes from an initial real constant value but remains positive for all times. The corresponding Lagrangian density is  
\begin{equation}
	\mathcal{L}=-\frac{1}{2}\bigl[\eta^{\mu\nu}\partial_{\mu}\phi\partial_{\nu}\phi+m^{2}(t)\phi^{2}\bigr]\,,
\end{equation}
with the Minkowski metric $\eta^{\mu\nu}$ taking the place of $\gamma^{\mu\nu}$ in \eqref{MetricPert}.  We use the convention with signature $(-1,+1,+1,+1)$. The corresponding equation of motion is
\begin{equation}
	-\partial_{\mu}\bigl(\eta^{\mu\nu}\partial_{\nu}\phi\bigr)+m^{2}\phi=0\,,
\end{equation}
Expanding the field variable $\phi$ in terms of the mode function $\phi_{\bm{k}}(t)$ of the form
\begin{align}\label{E:rhtbfdj}	
	\phi(\bm{x},t)&=\sum_{\bm{k}}e^{+i\bm{k}\cdot\bm{x}}\phi_{\bm{k}}(t)\,,&&\text{with}&\phi_{+\bm{k}}^{*}(t)&=\phi_{-\bm{k}}(t)\,.
\end{align}
we find the action given by
\begin{align}
	S=-\frac{1}{2}\!\int\!dt\sum_{\bm{k}}\Bigl[-\dot{\phi}_{\bm{k}}^{\vphantom{*}}\dot{\phi}_{\bm{k}}^{*}+\bigl(\bm{k}^{2}+m^{2}\bigr)\phi_{\bm{k}}^{\vphantom{*}}\phi_{\bm{k}}^{*}\Bigr]\,.\label{E:berjsf}
\end{align}
Since  $\dot{\phi}_{\bm{k}}^{\vphantom{*}}\dot{\phi}_{\bm{k}}^{*}=\dot{\phi}_{-\bm{k}}^{\vphantom{*}}\dot{\phi}_{-\bm{k}}^{*}$, and the summand in \eqref{E:berjsf} takes the same form for $\pm\bm{k}$,  we can write the action only in terms of the $\bm{k}>0$ modes
\begin{equation}\label{E:sohnfsdds}
	S=-\int\!dt\;\sum_{\bm{k}>0}\Bigl[-\dot{\phi}_{\bm{k}}^{\vphantom{*}}\dot{\phi}_{\bm{k}}^{*}+\bigl(\bm{k}^{2}+m^{2}\bigr)\phi_{\bm{k}}^{\vphantom{*}}\phi_{\bm{k}}^{*}\Bigr]\,,
\end{equation}

With the momentum conjugate to $\phi_{\bm{k}}$   defined by
\begin{equation}
	\pi_{\bm{k}}=\frac{\partial L}{\partial\dot{\phi}_{\bm{k}}}=\dot{\phi}_{\bm{k}}^{*}\,,
\end{equation}
 the Euler-Lagrange equation gives the equation of motion for  $\phi_{\bm{k}}(t)$
\begin{equation}\label{E:fgksgf}
	\ddot{\phi}_{\bm{k}}+\bigl(\bm{k}^{2}+m^{2}\bigr)\phi_{\bm{k}}=0\,.
\end{equation}

For the treatment of quantum fields we promote the field to be  operator. In the Heisenberg picture the expectation value will be computed with respect to the initial state. Then for each mode $\bm{k}$ we have
\begin{align}\label{E:kdfhsdf}
	\hat{\phi}_{+\bm{k}}^{\dagger}(t)&=\hat{\phi}_{-\bm{k}}^{\vphantom{\dagger}}(t)\,,&\hat{\pi}_{\bm{k}}&=\dot{\hat{\phi}}_{\bm{k}}^{\dagger}=\dot{\hat{\phi}}_{-\bm{k}}^{\vphantom{\dagger}}\,,
\end{align}
On the other hand, since the momentum $\hat{\pi}$ conjugate  to the field operator $\hat{\phi}$ is given by $\hat{\pi}=\dot{\hat{\phi}}$, we have
\begin{align}
	\hat{\pi}=\dot{\hat{\phi}}=\sum_{\bm{k}}\dot{\hat{\phi}}_{\bm{k}}\,e^{+i\bm{k}\cdot\bm{x}}=\sum_{\bm{k}}\dot{\hat{\phi}}_{-\bm{k}}^{\vphantom{\dagger}}\,e^{-i\bm{k}\cdot\bm{x}}=\sum_{\bm{k}}\dot{\hat{\phi}}_{\bm{k}}^{\dagger}\,e^{-i\bm{k}\cdot\bm{x}}=\sum_{\bm{k}}\hat{\pi}_{\bm{k}}\,e^{-i\bm{k}\cdot\bm{x}}\,.\label{E:qqzkgjbe}
\end{align}
The Fourier expansion of the momentum operator $\hat{\pi}$ takes a different form from that of $\hat{\phi}$ in \eqref{E:rhtbfdj}. This is   necessary  to ensure the standard form of the equal-time canonical commutation relation
\begin{equation}
	\bigl[\hat{\phi}(\bm{x},t),\hat{\pi}(\bm{x}',t)\bigr]=\sum_{\bm{k},\bm{k}'}\bigl[\hat{\phi}_{\bm{k}}(t),\hat{\pi}_{\bm{k}'}(t)\bigr]\,e^{i\bm{k}\cdot\bm{x}-i\bm{k}'\cdot\bm{x}'}=\sum_{\bm{k}}e^{i\bm{k}\cdot(\bm{x}-\bm{x}')}=i\,\delta^{(3)}(\bm{x}-\bm{x}')\,,
\end{equation}
if we require
\begin{equation}\label{E:eiubds}
	\bigl[\hat{\phi}_{\bm{k}}(t),\hat{\pi}_{\bm{k}'}(t)\bigr]=i\,\delta_{\bm{k},\bm{k}'}\,.
\end{equation}

In the general case, \eqref{E:kdfhsdf} allows a expansion
\begin{align}\label{E:rbgrivfghs}
	\hat{\phi}_{\bm{k}}(t)&=\hat{a}_{+\bm{k}}^{\vphantom{\dagger}}\,u_{\omega}^{\vphantom{*}}(t)+\hat{a}_{-\bm{k}}^{\dagger}\,u_{\omega}^{*}(t)\,,&&\text{and}&\hat{\pi}_{\bm{k}}(t)&=\hat{a}_{-\bm{k}}^{\vphantom{\dagger}}\,\dot{u}_{\omega}^{\vphantom{*}}(t)+\hat{a}_{+\bm{k}}^{\dagger}\,\dot{u}_{\omega}^{*}(t)\,.
\end{align}
Here $u_{\omega}(t)$ is a solution of \eqref{E:fgksgf}, satisfying the positive frequency condition at a specified initial time with $\omega^{2}(t)=\bm{k}^{2}+m^{2}(t)$, and $\hat{a}_{\bm{k}}^{\vphantom{\dagger}}$, $\hat{a}_{\bm{k}}^{\dagger}$ are the annihilation and creation operators associated with $u_{\omega}$ at the initial time, satisfying the usual commutation relations
\begin{align}\label{E:fgnrosd}
	\bigl[\hat{a}_{\bm{k}}^{\vphantom{\dagger}},\hat{a}_{\bm{k}'}^{\dagger}\bigr]&=\delta_{\bm{k}\bm{k}'}\,,
\end{align}
and zero otherwise. Then the commutation relation \eqref{E:eiubds} is consistently obeyed due to the Wronskian condition of $u_{\omega}$
\begin{equation}
	u_{\omega}^{\vphantom{*}}(t)\dot{u}_{\omega}^{*}(t)-u_{\omega}^{*}(t)\dot{u}_{\omega}^{\vphantom{*}}(t)=i\,.
\end{equation}
We may invert \eqref{E:rbgrivfghs} to express $\hat{a}_{\bm{k}}$ in terms of $\hat{\phi}_{\bm{k}}$ and $\hat{\pi}_{\bm{k}}$,
\begin{equation}
	\hat{a}_{\bm{k}}=-i\,\bigl(\dot{u}_{\omega}^{*}\,\hat{\phi}_{+\bm{k}}-u_{\omega}^{*}\,\hat{\pi}_{-\bm{k}}^{\vphantom{*}}\bigr)\,,
\end{equation}
and confirm the commutation relations \eqref{E:fgnrosd} for the creation and the annihilation operators.

From the expansions \eqref{E:rbgrivfghs}, we may compute the corresponding {correlation or} covariance matrix elements {among various modes}~\cite{SMD94}
\begin{align}
	\frac{1}{2}\langle\bigl\{\hat{\phi}_{\bm{k}},\,\hat{\phi}_{\bm{k}'}\bigr\}\rangle&=\frac{1}{2}\langle\bigl\{\hat{a}_{+\bm{k}}^{\vphantom{\dagger}},\hat{a}_{+\bm{k}'}^{\vphantom{\dagger}}\bigr\}\rangle\,u_{\omega}^{\vphantom{*}}u_{\omega'}^{\vphantom{*}}+\frac{1}{2}\langle\bigl\{\hat{a}_{-\bm{k}}^{\dagger},\hat{a}_{+\bm{k}'}^{\vphantom{\dagger}}\bigr\}\rangle\,u_{\omega}^{*}u_{\omega'}^{\vphantom{*}}+\frac{1}{2}\langle\bigl\{\hat{a}_{+\bm{k}}^{\vphantom{\dagger}},\hat{a}_{-\bm{k}'}^{\dagger}\bigr\}\rangle\,u_{\omega}^{\vphantom{*}}u_{\omega'}^{*}\notag\\
	&\qquad\qquad\qquad\qquad+\frac{1}{2}\langle\bigl\{\hat{a}_{-\bm{k}}^{\dagger},\hat{a}_{-\bm{k}'}^{\dagger}\bigr\}\rangle\,u_{\omega}^{*}u_{\omega'}^{*}\,,\label{E:fkgjhbdkg1}\\
	\frac{1}{2}\langle\bigl\{\hat{\phi}^{\vphantom{\dagger}}_{\bm{k}},\,\hat{\phi}^{\dagger}_{\bm{k}'}\bigr\}\rangle&=\frac{1}{2}\langle\bigl\{\hat{a}_{+\bm{k}}^{\vphantom{\dagger}},\hat{a}_{+\bm{k}'}^{\dagger}\bigr\}\rangle\,u_{\omega}^{\vphantom{*}}u_{\omega'}^{*}+\frac{1}{2}\langle\bigl\{\hat{a}_{-\bm{k}}^{\dagger},\hat{a}_{+\bm{k}'}^{\dagger}\bigr\}\rangle\,u_{\omega}^{*}u_{\omega'}^{*}+\frac{1}{2}\langle\bigl\{\hat{a}_{+\bm{k}}^{\vphantom{\dagger}},\hat{a}_{-\bm{k}'}^{\vphantom{\dagger}}\bigr\}\rangle\,u_{\omega}^{\vphantom{*}}u_{\omega'}^{\vphantom{*}}\notag\\
	&\qquad\qquad\qquad\qquad+\frac{1}{2}\langle\bigl\{\hat{a}_{-\bm{k}}^{\dagger},\hat{a}_{-\bm{k}'}^{\vphantom{\dagger}}\bigr\}\rangle\,u_{\omega}^{*}u_{\omega'}^{\vphantom{*}}\,,\\
	\frac{1}{2}\langle\bigl\{\hat{\phi}^{\vphantom{\dagger}}_{\bm{k}},\,\hat{\pi}^{\vphantom{\dagger}}_{\bm{k}'}\bigr\}\rangle&=\frac{1}{2}\langle\bigl\{\hat{a}_{+\bm{k}}^{\vphantom{\dagger}},\hat{a}_{+\bm{k}'}^{\dagger}\bigr\}\rangle\,u_{\omega}^{\vphantom{*}}\dot{u}_{\omega'}^{*}+\frac{1}{2}\langle\bigl\{\hat{a}_{-\bm{k}}^{\dagger},\hat{a}_{+\bm{k}'}^{\dagger}\bigr\}\rangle\,u_{\omega}^{*}\dot{u}_{\omega'}^{*}+\frac{1}{2}\langle\bigl\{\hat{a}_{+\bm{k}}^{\vphantom{\dagger}},\hat{a}_{-\bm{k}'}^{\vphantom{\dagger}}\bigr\}\rangle\,u_{\omega}^{\vphantom{*}}\dot{u}_{\omega'}^{\vphantom{*}}\notag\\
	&\qquad\qquad\qquad\qquad+\frac{1}{2}\langle\bigl\{\hat{a}_{-\bm{k}}^{\dagger},\hat{a}_{-\bm{k}'}^{\vphantom{\dagger}}\bigr\}\rangle\,u_{\omega}^{*}\dot{u}_{\omega'}^{\vphantom{*}}\,.\label{E:fkgjhbdkg3}
\end{align}
The results for $\langle\bigl\{\hat{\pi}_{\bm{k}},\,\hat{\pi}_{\bm{k}'}\bigr\}\rangle$ will be obtained by replacing $u_{\omega}$ in \eqref{E:fkgjhbdkg1} by $\dot{u}_{\omega}$. {These elements are extremely useful in constructing  physical observables of a Gaussian system in that, for each mode, we can simplify calculations involving an infinite-dimensional density matrix operator into those based a $2\times2$ covariance matrix.}

If the state used to evaluate the expectation value is a stationary state, i.e., $\langle\hat{a}_{\bm{k}}^{2}\rangle=0$, then \eqref{E:fkgjhbdkg1}--\eqref{E:fkgjhbdkg3} reduce to
\begin{align}
	\frac{1}{2}\langle\bigl\{\hat{\phi}_{\bm{k}},\,\hat{\phi}_{\bm{k}'}\bigr\}\rangle&=\delta_{+\bm{k},-\bm{k}'}\Bigl[\Bigl(\langle\hat{N}_{+\bm{k}}\rangle+\frac{1}{2}\Bigr)+\Bigl(\langle\hat{N}_{-\bm{k}}\rangle+\frac{1}{2}\Bigr)\Bigr]\,u_{\omega}^{\vphantom{*}}u_{\omega}^{*}\,,\label{E:ngsjggh1}\\
	\frac{1}{2}\langle\bigl\{\hat{\phi}^{\vphantom{\dagger}}_{\bm{k}},\,\hat{\phi}^{\dagger}_{\bm{k}'}\bigr\}\rangle&=\delta_{+\bm{k},+\bm{k}'}\Bigl[\Bigl(\langle\hat{N}_{+\bm{k}}\rangle+\frac{1}{2}\Bigr)+\Bigl(\langle\hat{N}_{-\bm{k}}\rangle+\frac{1}{2}\Bigr)\Bigr]\,u_{\omega}^{\vphantom{*}}u_{\omega}^{*}\,,\\
	\frac{1}{2}\langle\bigl\{\hat{\phi}^{\vphantom{\dagger}}_{\bm{k}},\,\hat{\pi}^{\vphantom{\dagger}}_{\bm{k}'}\bigr\}\rangle&=\delta_{+\bm{k},+\bm{k}'}\Bigl[\Bigl(\langle\hat{N}_{+\bm{k}}\rangle+\frac{1}{2}\Bigr)u_{\omega}^{\vphantom{*}}\dot{u}_{\omega}^{*}+\Bigl(\langle\hat{N}_{-\bm{k}}\rangle+\frac{1}{2}\Bigr)\dot{u}_{\omega}^{\vphantom{*}}u_{\omega}^{*}\Bigr]\,.\label{E:ngsjggh3}
\end{align}
where $\hat{N}_{\bm{k}}=\hat{a}_{\bm{k}}^{\dagger}\hat{a}_{\bm{k}}^{\vphantom{\dagger}}$ is the number operator of mode $\bm{k}$. Here we see that both $\pm\bm{k}$ modes contribute. Eqs.~\eqref{E:ngsjggh1}--\eqref{E:ngsjggh3} imply that 
\begin{align}
	\langle\hat{\phi}_{\pm\bm{k}}^{2}(t)\rangle&=0\,,&\langle\hat{\phi}_{\pm\bm{k}}^{\vphantom{\dagger}}(t)\hat{\pi}_{\pm\bm{k}}^{\dagger}(t)\rangle&=0\,.
\end{align}
These ingredients will allow us to construct the density matrix operator of the field, which will be used to compute the expectation values or observables of the field.

\section{Nonequilibrium Evolution of the Density Matrix}\label{S:ebsfg}

We now proceed to construct the density matrix operator using  the formalisms developed in \cite{NEqFE}. We begin by writing the field mode operators in the form of harmonic oscillators.

As the operators $\hat{\phi}_{\bm{k}}$, $\hat{\pi}_{\bm{k}}$  are not Hermitian,  we decompose them into the real and the imaginary parts by
\begin{align}\label{E:gkjd1}
	\hat{\phi}_{\bm{k}}&=\frac{1}{\sqrt{2}}\,\bigl(\hat{\phi}_{\bm{k}}^{(1)}+i\,\hat{\phi}_{\bm{k}}^{(2)}\bigr)\,.
\end{align}
Hence the action \eqref{E:sohnfsdds} becomes
\begin{align}\label{E:fkbdfrert}
	S=\frac{1}{2}\int\!dt\;\sum_{\bm{k}>0}\Bigl[\Bigl(\dot{\phi}_{\bm{k}}^{(1)2}+\dot{\phi}_{\bm{k}}^{(2)2}\Bigr)-\omega^{2}(t)\Bigl(\phi_{\bm{k}}^{(1)2}+\phi_{\bm{k}}^{(2)2}\Bigr)\Bigr]\,.
\end{align}
with the real and the imaginary parts acting as uncoupled parametric oscillators.

The corresponding conjugate momenta are  given by
\begin{align}
	p_{\bm{k}}^{(1)}&=\frac{\partial L}{\partial\dot{\phi}_{\bm{k}}^{(1)}}=\dot{\phi}_{\bm{k}}^{(1)}\,,&p_{\bm{k}}^{(2)}&=\frac{\partial L}{\partial\dot{\phi}_{\bm{k}}^{(2)}}=\dot{\phi}_{\bm{k}}^{(2)}\,,
\end{align}
which implies
\begin{equation}\label{E:gkjd2}
	\hat{\pi}_{\bm{k}}=\dot{\hat{\phi}}_{\bm{k}}^{\dagger}=\frac{1}{\sqrt{2}}\bigl(p_{\bm{k}}^{(1)}-i\,p_{\bm{k}}^{(2)}\bigr)\,.
\end{equation}
Note the sign before the imaginary part of $\hat{\pi}_{\bm{k}}$. This then ensures that Eq.~\eqref{E:eiubds} is obeyed
\begin{equation}
	\bigl[\hat{\phi}_{\bm{k}},\hat{\pi}_{\bm{k}'}\bigr]=\frac{1}{2}\bigl[\hat{\phi}_{\bm{k}}^{(1)}+i\,\hat{\phi}_{\bm{k}}^{(2)},\,p_{\bm{k}}^{(1)}-i\,p_{\bm{k}}^{(2)}\bigr]=i\,\delta_{\bm{k},\bm{k}'}\,,
\end{equation}
if $\bigl[\hat{\phi}_{\bm{k}}^{(i)},\,p_{\bm{k}'}^{(j)}\bigr]=i\,\delta_{ij}\delta_{\bm{k},\bm{k}'}$. Furthermore since $\hat{\phi}_{+\bm{k}}^{\dagger}=\hat{\phi}_{-\bm{k}}^{\vphantom{\dagger}}$, we find
\begin{align}
	\hat{\phi}_{+\bm{k}}^{(1)}&=+\hat{\phi}_{-\bm{k}}^{(1)}\,,&&\text{and}&\hat{\phi}_{+\bm{k}}^{(2)}&=-\hat{\phi}_{-\bm{k}}^{(2)}\,.
\end{align}
That is, $\hat{\phi}_{\bm{k}}^{(1)}$ is an even function of $\bm{k}$, but $\hat{\phi}_{\bm{k}}^{(2)}$ is an odd function of $\bm{k}$.

To construct the density matrix operator, we need the covariance matrix elements of $\hat{\phi}_{\bm{k}}^{(i)}$. This can be done by first inverting the decompositions \eqref{E:gkjd1} and \eqref{E:gkjd2}
\begin{align}
	\hat{\phi}_{\bm{k}}^{(1)}&=\frac{1}{\sqrt{2}}\bigl(\hat{\phi}_{\bm{k}}^{\vphantom{\dagger}}+\hat{\phi}_{\bm{k}}^{\dagger}\bigr)\,,&\hat{\phi}_{\bm{k}}^{(2)}&=\frac{1}{i\sqrt{2}}\bigl(\hat{\phi}_{\bm{k}}^{\vphantom{\dagger}}-\hat{\phi}_{\bm{k}}^{\dagger}\bigr)\,,\label{E:rkjgb1}\\
	\hat{p}_{\bm{k}}^{(1)}&=\frac{1}{\sqrt{2}}\bigl(\hat{\pi}_{\bm{k}}^{\dagger}+\hat{\pi}_{\bm{k}}^{\vphantom{\dagger}}\bigr)\,,&\hat{p}_{\bm{k}}^{(2)}&=\frac{1}{i\sqrt{2}}\bigl(\hat{\pi}_{\bm{k}}^{\dagger}-\hat{\phi}_{\bm{k}}^{\vphantom{\dagger}}\bigr)\,.\label{E:rkjgb2}
\end{align}
From \eqref{E:ngsjggh1}--\eqref{E:ngsjggh3}, we obtain the covariance matrix elements of $\hat{\phi}_{\bm{k}}^{(i)}$ with respect to the initial stationary state given by
\begin{align}
	\langle\hat{\phi}_{\bm{k}}^{(1)2}\rangle&=\Bigl[\Bigl(\langle\hat{N}_{+\bm{k}}\rangle+\frac{1}{2}\Bigr)+\Bigl(\langle\hat{N}_{-\bm{k}}\rangle+\frac{1}{2}\Bigr)\Bigr]\,u_{\omega}^{\vphantom{*}}u_{\omega}^{*}=\langle\hat{\phi}_{-\bm{k}}^{(1)2}\rangle\,,\label{E:fkgbte1}\\
	\frac{1}{2}\langle\bigl\{\hat{\phi}_{\bm{k}}^{(1)},\,\hat{p}_{\bm{k}}^{(1)}\bigr\}\rangle&=\frac{1}{2}\Bigl[\Bigl(\langle\hat{N}_{+\bm{k}}\rangle+\frac{1}{2}\Bigr)+\Bigl(\langle\hat{N}_{-\bm{k}}\rangle+\frac{1}{2}\Bigr)\Bigr]\Bigl(u_{\omega}^{\vphantom{*}}\dot{u}_{\omega}^{*}+\dot{u}_{\omega}^{\vphantom{*}}u_{\omega}^{*}\Bigr)=\frac{1}{2}\langle\bigl\{\hat{\phi}_{-\bm{k}}^{(1)},\,\hat{p}_{-\bm{k}}^{(1)}\bigr\}\rangle\,,\label{E:fkgbte2}
\end{align}
and then
\begin{align}
	\langle\hat{\phi}_{\bm{k}}^{(2)2}\rangle&=\langle\hat{\phi}_{\bm{k}}^{(1)2}\rangle\,,&\frac{1}{2}\langle\bigl\{\hat{\phi}_{\bm{k}}^{(2)},\,\hat{p}_{\bm{k}}^{(2)}\bigr\}\rangle&=\frac{1}{2}\langle\bigl\{\hat{\phi}_{\bm{k}}^{(1)},\,\hat{p}_{\bm{k}}^{(1)}\bigr\}\rangle\,.\label{E:fkgbte3}
\end{align}
Finally, we find
\begin{align}\label{E:fkgbte4}
	\frac{1}{2}\langle\bigl\{\hat{\phi}_{\bm{k}}^{(1)},\,\hat{\phi}_{\bm{k}}^{(2)}\bigr\}\rangle&=0\,,&\frac{1}{2}\langle\bigl\{\hat{\phi}_{\bm{k}}^{(1)},\,\hat{p}_{\bm{k}}^{(2)}\bigr\}\rangle&=0
\end{align}
for the initial stationary state. 

Having seen how the field modes are expressed in terms of  harmonic oscillators, we can now construct the density matrix operators of the field modes.

\subsection{Density matrix operator}

Let us quickly review  the general form of the density matrix operator of a Gaussian state~\cite{NEqFE}. Consider the one-mode case, and let $(\hat{q},\hat{p})$ be the conjugated canonical operators of a Gaussian system. The density operator of the Gaussian state in general takes the form
\begin{align}
	\hat{\rho}(\hat{q},\hat{p})=\frac{2}{\sqrt{e^{2\varphi}-1}}\,\exp\biggl\{-e^{-\varphi}\cosh^{-1}\coth\varphi\,\Bigl[\mathsf{a}\,\hat{q}^{2}+\mathsf{b}\,\hat{p}^{2}-\mathsf{c}\,\bigl\{\hat{q},\,\hat{p}\bigr\}\Bigr]\biggr\}\,,\label{E:lrngflgs}
\end{align}
where the coefficients can be expressed in terms of the corresponding time-dependent covariance matrix elements
\begin{align}
	\mathsf{a}&=\langle\hat{p}^{2}\rangle\,,&\mathsf{b}&=\langle\hat{q}^{2}\rangle\,,&\mathsf{c}&=\frac{1}{2}\langle\bigl\{\hat{q},\hat{p}\bigr\}\rangle\,,&&\text{and}&e^{2\varphi}&=4\bigl(\mathsf{a}\mathsf{b}-\mathsf{c}^{2}\bigr)\,.
\end{align}
By means of the creation and annihilation operators, the density matrix operator \eqref{E:lrngflgs} becomes
\begin{align}
	\hat{\rho}(\hat{a},\hat{a}^{\dagger})=\frac{2}{\sqrt{e^{2\varphi}-1}}\,\exp\biggl\{-2\Xi\,e^{-\varphi}\cosh^{-1}\coth\varphi\,\Bigl[\kappa\,\hat{a}^{2}+\kappa^{*}\,\hat{a}^{\dagger2}+\frac{\tau}{2}\,\bigl\{\hat{a},\,\hat{a}^{\dagger}\bigr\}\Bigr]\biggr\}\,,\label{E:lgkndfgds}
\end{align}
where
\begin{align}
	\kappa&=\frac{1}{2}\,\sinh2\eta\,e^{-i\psi}\,,&\tau&=\cosh2\eta\,,\label{E:kgbskf1}\\
	\Xi&=\coth\frac{\beta}{2}\,,&e^{2\varphi}&=\Xi^{2}\bigl(\tau^{2}-4\lvert\kappa\rvert^{2}\bigr)=\Xi^{2}=\coth^{2}\frac{\beta}{2}\,,\label{E:kgbskf2}
\end{align}
and $\beta$, $\zeta=\eta\,e^{i\psi}$ are inverse-temperature-like and squeeze-like parameters. Hence we have
\begin{align}\label{E:fbgdjhbf}
	2\Xi\,e^{-\varphi}\cosh^{-1}\coth\varphi&=2\beta\,,&\frac{2}{\sqrt{e^{2\varphi}-1}}&=2\sinh\frac{\beta}{2}\,.
\end{align}
The latter is essentially the inverse of the nonequilibrium partition function~\cite{NEqFE} associated with the density matrix operator \eqref{E:lrngflgs} or \eqref{E:lgkndfgds}. We have assumed that the first moments vanish for the state considered; otherwise, we may let $\hat{a}\to\hat{a}-\langle\hat{a}\rangle$ or $\hat{q}\to\hat{q}-\langle\hat{q}\rangle$ for the more general cases.

Observe that  the factor
\begin{equation}\label{E:dfkbkrt}
	\kappa\,\hat{a}^{2}+\kappa^{*}\,\hat{a}^{\dagger2}+\frac{\tau}{2}\,\bigl\{\hat{a},\,\hat{a}^{\dagger}\bigr\}
\end{equation}
can be `rotated' into the form
\begin{equation}
	\frac{1}{2}\,\bigl\{\hat{b},\,\hat{b}^{\dagger}\bigr\}\,.
\end{equation}
Suppose such a $\hat{b}$ is related to $\hat{a}$ and $\hat{a}^{\dagger}$ by $\hat{b}=\mu\,\hat{a}+\nu\,\hat{a}^{\dagger}$, with $\lvert\mu\rvert^{2}-\lvert\nu\rvert^{2}=1$,   we have
\begin{align}
	\frac{1}{2}\,\bigl\{\hat{b},\,\hat{b}^{\dagger}\bigr\}&=\frac{\mu\nu^{*}}{2}\,\bigl\{\hat{a},\hat{a}\bigr\}+\frac{\mu^{*}\nu}{2}\,\bigl\{\hat{a}^{\dagger},\hat{a}^{\dagger}\bigr\}+\frac{\lvert\mu\rvert^{2}+\lvert\nu\rvert^{2}}{2}\,\bigl\{\hat{a},\hat{a}^{\dagger}\bigr\}\,.
\end{align}
Comparing with \eqref{E:dfkbkrt}, we obtain
\begin{align}
	\mu\nu^{*}&=\kappa=\frac{1}{2}\,\sinh2\eta\,e^{-i\psi}\,,&\lvert\mu\rvert^{2}+\lvert\nu\rvert^{2}&=2\bigl(\lvert\nu\rvert^{2}+\frac{1}{2}\bigr)=\tau=\cosh2\eta\,.
\end{align}
If we assume $\mu\in\mathbb{R}$, then a useful choice of $(\mu,\nu)$ is
\begin{align}\label{E:gfkbsfg}
	\mu&=\cosh\eta\,,&\nu&=\sinh\eta\,e^{-i\psi}\,,
\end{align}
that is, a squeeze transformation. A complex $\mu$ will correspond to an additional rotation, so that \eqref{E:gfkbsfg} will contain the phase factor that accounts for the rotation.

Since from \eqref{E:fkbdfrert}, the real part and the imaginary parts of $\hat{\phi}_{\bm{k}}$ behave  like independent harmonic oscillators, we expect that the corresponding  density operator for each $\bm{k}>0$ mode will be
\begin{align}\label{E:dgkhs}
	\hat{\rho}(\hat{\phi}^{(1)}_{\bm{k}},\hat{\phi}^{(2)}_{\bm{k}})&=\biggl(\frac{2}{\sqrt{e^{2\varphi_{\bm{k}}}-1}}\biggr)^{2}\,\exp\biggl\{-e^{-\varphi_{\bm{k}}}\cosh^{-1}\coth\varphi_{\bm{k}}\,\Bigl[\mathsf{a}_{\bm{k}}^{(1)}\hat{\phi}_{\bm{k}}^{(1)2}+\mathsf{b}_{\bm{k}}^{(1)}\hat{p}_{\bm{k}}^{(1)2}-\mathsf{c}_{\bm{k}}^{(1)}\bigl\{\hat{\phi}_{\bm{k}}^{(1)},\,\hat{p}_{\bm{k}}^{(1)}\bigr\}\Bigr.\biggr.\notag\\
	&\qquad\qquad\qquad\qquad\qquad\qquad\qquad+\biggl.\Bigl.\mathsf{a}_{\bm{k}}^{(2)}\hat{\phi}_{\bm{k}}^{(2)2}+\mathsf{b}_{\bm{k}}^{(2)}\hat{p}_{\bm{k}}^{(2)2}-\mathsf{c}_{\bm{k}}^{(2)}\bigl\{\hat{\phi}_{\bm{k}}^{(2)},\,\hat{p}_{\bm{k}}^{(2)}\bigr\}\Bigr]\biggr\}\,,
\end{align}
where we have assumed that $\varphi_{\bm{k}}$ is the same for the real and the imaginary components, to be justified later. Eqs.~\eqref{E:fkgbte1}--\eqref{E:fkgbte4} imply that
\begin{align}
	\mathsf{a}_{+\bm{k}}^{(2)}&=\mathsf{a}_{+\bm{k}}^{(1)}=\mathsf{a}_{-\bm{k}}^{(1)}=\mathsf{a}_{-\bm{k}}^{(2)}\,,&\mathsf{b}_{+\bm{k}}^{(2)}&=\mathsf{b}_{+\bm{k}}^{(1)}=\mathsf{b}_{-\bm{k}}^{(1)}=\mathsf{b}_{-\bm{k}}^{(2)}\,,&\mathsf{c}_{+\bm{k}}^{(2)}&=\mathsf{c}_{+\bm{k}}^{(1)}=\mathsf{c}_{-\bm{k}}^{(1)}=\mathsf{c}_{-\bm{k}}^{(2)}\,,\label{E:eirh1}
\end{align}
so let us assign
\begin{align}
	\mathsf{a}_{\bm{k}}&\equiv \mathsf{a}_{\bm{k}}^{(1)}=\mathsf{a}_{\bm{k}}^{(2)}\,,&\mathsf{b}_{\bm{k}}&\equiv \mathsf{b}_{\bm{k}}^{(1)}=\mathsf{b}_{\bm{k}}^{(2)}\,,&\mathsf{c}_{\bm{k}}&\equiv \mathsf{c}_{\bm{k}}^{(1)}=\mathsf{c}_{\bm{k}}^{(2)}\,,\label{E:eirh2}
\end{align}
and write Eq.~\eqref{E:dgkhs} as
\begin{align}
	\hat{\rho}(\hat{\phi}^{(1)}_{\bm{k}},\hat{\phi}^{(2)}_{\bm{k}})&=\biggl(\frac{2}{\sqrt{e^{2\varphi_{\bm{k}}}-1}}\biggr)^{2}\,\exp\biggl\{-e^{-\varphi_{\bm{k}}}\cosh^{-1}\coth\varphi_{\bm{k}}\,\Bigl[\mathsf{a}_{\bm{k}}\Bigl(\hat{\phi}_{\bm{k}}^{(1)2}+\hat{\phi}_{\bm{k}}^{(2)2}\Bigr)+\mathsf{b}_{\bm{k}}\Bigl(\hat{p}_{\bm{k}}^{(1)2}+\hat{p}_{\bm{k}}^{(2)2}\Bigr)\Bigr.\biggr.\notag\\
	&\qquad\qquad\qquad\qquad\qquad\qquad\qquad-\biggl.\Bigl.\mathsf{c}_{\bm{k}}\Bigl(\bigl\{\hat{\phi}_{\bm{k}}^{(1)},\,\hat{p}_{\bm{k}}^{(1)}\bigr\}+\bigl\{\hat{\phi}_{\bm{k}}^{(2)},\,\hat{p}_{\bm{k}}^{(2)}\bigr\}\Bigr)\Bigr]\biggr\}\,,\label{E:gksgbsd}
\end{align}
with $\bm{k}>0$.

When we take into account all modes, the density matrix operator of the field modes $\hat{\rho}(\{\hat{\phi}_{\bm{k}}\},\{\hat{\pi}_{\bm{k}}\})$ corresponding to \eqref{E:fkbdfrert} can be found to be
\begin{align}\label{E:ngsjgfd}
	\hat{\rho}(\{\hat{\phi}_{\bm{k}}\},\{\hat{\pi}_{\bm{k}}\})&=\prod_{\bm{k}>0}\hat{\rho}(\hat{\phi}^{(1)}_{\bm{k}},\hat{\phi}^{(2)}_{\bm{k}})
\\	
&=\biggl(\prod_{\bm{k}}\frac{2}{\sqrt{e^{2\varphi_{\bm{k}}}-1}}\biggr)\exp\biggl\{-\sum_{\bm{k}}e^{-\varphi_{\bm{k}}}\cosh^{-1}\coth\varphi_{\bm{k}}\Bigl[\mathsf{a}_{\bm{k}}\,\hat{\phi}_{+\bm{k}}^{\vphantom{\dagger}}\hat{\phi}_{+\bm{k}}^{\dagger}+\mathsf{b}_{\bm{k}}\,\hat{\pi}_{+\bm{k}}^{\vphantom{\dagger}}\hat{\pi}_{+\bm{k}}^{\dagger}-\mathsf{c}_{\bm{k}}\,\bigl\{\hat{\phi}_{\bm{k}},\,\hat{\pi}_{\bm{k}}\bigr\}\Bigr]\biggr\}\,,\notag
\end{align}
where we have assumed that the parameter $\varphi_{+\bm{k}}=\varphi_{-\bm{k}}$, to be justified later.  What remains is to  express the coefficients $\mathsf{b}_{\bm{k}}$, $\mathsf{a}_{\bm{k}}$ and $\mathsf{c}_{\bm{k}}$ by the original field mode operators $\hat{\phi}_{\bm{k}}$ and $\hat{\pi}_{\bm{k}}$. Since $\mathsf{b}_{\bm{k}}=\langle\hat{\phi}_{\bm{k}}^{(1)2}\rangle=\langle\hat{\phi}_{\bm{k}}^{(2)2}\rangle$, we find
\begin{align}
	\mathsf{b}_{\bm{k}}=\frac{1}{2}\bigl(\langle\hat{\phi}_{\bm{k}}^{(1)2}\rangle+\langle\hat{\phi}_{\bm{k}}^{(2)2}\rangle\bigr)=\frac{1}{2}\bigl(\langle\hat{\phi}_{\bm{k}}^{\vphantom{\dagger}}\hat{\phi}_{\bm{k}}^{\dagger}\rangle+\langle\hat{\phi}_{-\bm{k}}^{\vphantom{\dagger}}\hat{\phi}_{-\bm{k}}^{\dagger}\rangle\bigr)=\frac{1}{2}\langle\bigl\{\hat{\phi}_{\bm{k}}^{\vphantom{\dagger}},\,\hat{\phi}_{\bm{k}}^{\dagger}\bigr\}\rangle\,.
\end{align}
Likewise, from
\begin{align}
	\mathsf{c}_{\bm{k}}=\frac{1}{2}\langle\bigl\{\hat{\phi}_{\bm{k}}^{(1)},\,\hat{p}_{\bm{k}}^{(1)}\bigr\}\rangle=\frac{1}{2}\langle\bigl\{\hat{\phi}_{\bm{k}}^{(2)},\,\hat{p}_{\bm{k}}^{(2)}\bigr\}\rangle\,,
\end{align}
we obtain
\begin{align}
	\mathsf{c}_{\bm{k}}=\frac{1}{2}\Bigl[\frac{1}{2}\langle\bigl\{\hat{\phi}_{\bm{k}},\,\hat{\pi}_{\bm{k}}\bigr\}\rangle+\frac{1}{2}\langle\bigl\{\hat{\phi}_{\bm{k}}^{\dagger},\,\hat{\pi}_{\bm{k}}^{\dagger}\bigr\}\rangle\Bigr]=\frac{1}{2}\,\Bigl[\frac{1}{2}\langle\bigl\{\hat{\phi}_{\bm{k}},\,\hat{\pi}_{\bm{k}}\bigr\}\rangle+\frac{1}{2}\langle\bigl\{\hat{\phi}_{-\bm{k}},\,\hat{\pi}_{-\bm{k}}\bigr\}\rangle\Bigr]\,.
\end{align}
At the first sight, the result for $\mathsf{c}_{\bm{k}}$ is not quite satisfactory because we cannot clearly separate contributions from the $\pm\bm{k}$ modes. However, later we will see this is the form  needed to construct the Wigner function of the whole field, instead of for each mode. This shows explicitly that there is an inseparable correlation between the $\pm\bm{k}$ modes.

In terms of the mode expansion \eqref{E:rbgrivfghs}, the elements $\mathsf{b}_{\bm{k}}$, $\mathsf{a}_{\bm{k}}$ and $\mathsf{c}_{\bm{k}}$ take the forms
\begin{align}
	\mathsf{b}_{\bm{k}}&=\Bigl[\Bigl(\langle\hat{N}_{+\bm{k}}\rangle+\frac{1}{2}\Bigr)+\Bigl(\langle\hat{N}_{-\bm{k}}\rangle+\frac{1}{2}\Bigr)\Bigr]\,u_{\omega}^{\vphantom{*}}u_{\omega}^{*}\,,\label{E:rhee1}\\
	\mathsf{a}_{\bm{k}}&=\Bigl[\Bigl(\langle\hat{N}_{+\bm{k}}\rangle+\frac{1}{2}\Bigr)+\Bigl(\langle\hat{N}_{-\bm{k}}\rangle+\frac{1}{2}\Bigr)\Bigr]\,\dot{u}_{\omega}^{\vphantom{*}}\dot{u}_{\omega}^{*}\,,\\
	\mathsf{c}_{\bm{k}}&=\frac{1}{2}\Bigl[\Bigl(\langle\hat{N}_{+\bm{k}}\rangle+\frac{1}{2}\Bigr)+\Bigl(\langle\hat{N}_{-\bm{k}}\rangle+\frac{1}{2}\Bigr)\Bigr]\bigl(u_{\omega}^{\vphantom{*}}\dot{u}_{\omega}^{*}+\dot{u}_{\omega}^{\vphantom{*}}u_{\omega}^{*}\bigr)\,.\label{E:rhee3}
\end{align}
with $\hat{N}_{\bm{k}}^{\vphantom{\dagger}}=\hat{a}_{\bm{k}}^{\dagger}\hat{a}_{\bm{k}}^{\vphantom{\dagger}}$, if the state is stationary. We again observe that they are the same for $\pm\bm{k}$ modes and thus the corresponding parameters $\varphi_{\bm{k}}$, $\beta_{\bm{k}}$, $\zeta_{\bm{k}}$, $\kappa_{\bm{k}}$ and $\tau_{\bm{k}}$ in \eqref{E:kgbskf1} and \eqref{E:kgbskf2} will also be  the same for $\pm\bm{k}$ modes,  because they are functions of the covariance matrix elements $\mathsf{b}_{\bm{k}}$, $\mathsf{a}_{\bm{k}}$ and $\mathsf{c}_{\bm{k}}$. Note that from~\cite{NEqFE}, we observe that the density matrix operator thus constructed is fairly general for the configuration under study.  {In addition, it is worth  noticing that since the second expression in \eqref{E:fbgdjhbf} can be identified as the inverse of the nonequilibrium partition function for the canonical variables $(\hat{\phi}_{\bm{k}},\hat{\pi}_{\bm{k}})$, we find that the prefactor before the exponential in \eqref{E:ngsjgfd} turns out to be the inverse of the nonequilibrium partition function of the quantum field, from which we can extract the nonequilibrium free energy of the field. Since these quantities are nonequilibrium in nature, they are all functions of time. Moreover, since the field is driven by the parametric process, the state of the field at any moment is a very general Gaussian state. It means that the nonequilibrium partition function or the nonequilibrium free energy for such a Gaussian system is introduced in a context much more general than conventional equilibrium quantum thermodynamics.}

\subsection{Density matrix elements}

The  {matrix elements of the density operator of} a Gaussian state in the $\phi_{\bm{k}}^{(i)}$ basis can be written as
\begin{align}\label{E:gdxkjs}
	\varrho(\phi^{(i)}_{\bm{k}},\phi'^{(i)}_{\bm{k}})&=\mathcal{N}_{\varrho}\exp\Bigl(-\mathcal{A}_{\bm{k}}^{\vphantom{2}}\Delta_{\bm{k}}^{(i)2}-i\,2\mathcal{B}_{\bm{k}}^{\vphantom{2}}\Delta_{\bm{k}}^{(i)}\Sigma_{\bm{k}}^{(i)}-\mathcal{C}_{\bm{k}}^{\vphantom{2}}\Sigma_{\bm{k}}^{(i)2}\Bigr)\,,
\end{align}
where $i=1$, $2$, and the relative coordinate $\Delta_{\bm{k}}^{(i)}$, center-of-mass coordinate $\Sigma_{\bm{k}}^{(i)}$, and normalization constant $\mathcal{N}_{\varrho}$ are respectively given by
\begin{align}
	\Delta_{\bm{k}}^{(i)}&=\phi^{(i)}_{\bm{k}}-\phi'^{(i)}_{\bm{k}}\,,&\Sigma_{\bm{k}}^{(i)}&=\frac{\phi^{(i)}_{\bm{k}}+\phi'^{(i)}_{\bm{k}}}{2}\,,&\mathcal{N}_{\varrho}&=\frac{\sqrt{\mathcal{C}_{\bm{k}}}}{\sqrt{\pi}}\,,
\end{align}
and
\begin{align}
	\mathcal{A}_{\bm{k}}&=\frac{\mathsf{a}_{\bm{k}}^{\vphantom{2}}\mathsf{b}_{\bm{k}}^{\vphantom{2}}-\mathsf{c}_{\bm{k}}^{2}}{2\mathsf{b}_{\bm{k}}}\,,&\mathcal{B}_{\bm{k}}&=-\frac{\mathsf{c}_{\bm{k}}}{2\mathsf{b}_{\bm{k}}}\,,&\mathcal{C}_{\bm{k}}&=\frac{1}{2\mathsf{b}_{\bm{k}}}\,,
\end{align}
with
\begin{align}
	\mathsf{b}_{\bm{k}}&=\langle\phi_{\bm{k}}^{(i)2}\rangle\,,&\mathsf{a}_{\bm{k}}&=\langle p_{\bm{k}}^{(i)2}\rangle\,,&\mathsf{c}_{\bm{k}}&=\frac{1}{2}\langle\bigl\{\phi_{\bm{k}}^{(i)},p_{\bm{k}}^{(i)}\bigr\}\rangle\,.
\end{align}
If the initial state of the density matrix is a vacuum state, then we always have
\begin{equation}
	\mathsf{b}_{\bm{k}}^{\vphantom{2}}\mathsf{a}_{\bm{k}}^{\vphantom{2}}-\mathsf{c}_{\bm{k}}^{2}=\frac{1}{4}\,,
\end{equation}
so that \eqref{E:gdxkjs} reduces to
\begin{align}\label{E:hhweh}
	\varrho(\phi^{(i)}_{\bm{k}},\phi'^{(i)}_{\bm{k}})&=\mathcal{N}_{\varrho}\exp\Bigl[-\frac{1}{4b_{\bm{k}}}\Bigl(\phi_{\bm{k}}^{(i)2}+\phi'^{(i)2}_{\bm{k}}\Bigr)+i\,\frac{c_{\bm{k}}}{2b_{\bm{k}}}\Bigl(\phi_{\bm{k}}^{(i)2}-\phi'^{(i)2}_{\bm{k}}\Bigr)\Bigr]\,.
\end{align}

\section{Evolution of mode functions: Bogoliubov Transformation}\label{S:ebgf}

The dynamics of the field is fully encoded in its density matrix operator or its elements. The time evolution of the density matrix can be described by the unitary  time-evolution operator $\hat{U}(t_{f},t_{i})$
\begin{equation}
	\hat{\rho}(t_{f})=\hat{U}(t_{f},t_{i})\,\hat{\rho}(t_{i})\,\hat{U}^{\dagger}(t_{f},t_{i})\,.
\end{equation}
A more succinct description of the evolution of a linear quantum system is by using the language of  Bogoliubov transformations in the Heisenberg picture.  We begin with a brief overview of the Bogoliubov transformation.

\subsection{Active and passive views of the Bogoliubov transformation}
Suppose  the quantum scalar field $\hat{\phi}$ is expanded with respect to two different sets of mode functions $u_{\bm{k}}^{(\textsc{in})}(t)$ and $u_{\bm{k}}^{(\textsc{out})}(t)$. Then according to \eqref{E:rbgrivfghs} the expansion takes the form
\begin{align}
	\hat{\phi}(\bm{x},t)&=\sum_{\bm{k}}\Bigl[\hat{a}_{\bm{k}}^{\vphantom{\dagger}}\,u_{\bm{k}}^{(\textsc{in})}(\bm{x},t)+\hat{a}_{\bm{k}}^{\dagger}\,u_{\bm{k}}^{(\textsc{in})*}(\bm{x},t)\Bigr]\,,
	\intertext{or,  in terms of the out-mode functions, in the form}
	\hat{\phi}(\bm{x},t)&=\sum_{\bm{k}}\Bigl[\hat{b}_{\bm{k}}^{\vphantom{\dagger}}\,u_{\bm{k}}^{(\textsc{out})}(\bm{x},t)+\hat{b}_{\bm{k}}^{\dagger}\,u_{\bm{k}}^{(\textsc{out})*}(\bm{x},t)\Bigr]\,,
\end{align}
where $\hat{a}_{\bm{k}}$ and $\hat{b}_{\bm{k}}$ are the annihilation operators associated $u_{\bm{k}}^{(\textsc{in})}(t)$ and $u_{\bm{k}}^{(\textsc{out})}(t)$ respectively. Both sets are assumed to be related by a transformation
\begin{align}\label{E:rthdnf1}
	u_{\bm{k}}^{(\textsc{in})}(\bm{x},t)&=\alpha_{\bm{k}}^{\vphantom{(\textsc{out})}}u_{\bm{k}}^{(\textsc{out})}(\bm{x},t)+\beta_{\bm{k}}^{\vphantom{(\textsc{out})}}u_{-\bm{k}}^{(\textsc{out})*}(\bm{x},t)\,,&&\text{with}&u_{\bm{k}}^{(\textsc{in})}(\bm{x},t)&=e^{i\bm{k}\cdot\bm{x}}u_{\omega}^{(\textsc{in})}(t)\,,
\end{align}
where $\alpha_{\bm{k}}$, $\beta_{\bm{k}}\in\mathbb{C}$ are called the Bogoliubov coefficients~\cite{InfSqV,FDRSq,LF21}, satisfying 
\begin{align}\label{E:rubses}
	\lvert\alpha_{\bm{k}}\rvert^{2}-\lvert\beta_{\bm{k}}\rvert^{2}=1\,,
\end{align}
for the configuration of our interest, and $u_{\omega}^{(\textsc{in})}(t)$ is the correspondence of $u_{\omega}(t)$ in \eqref{E:rbgrivfghs} for the IN modes. In fact, Eq.~\eqref{E:rthdnf1} is equivalent to
\begin{equation}\label{E:bgskgsd}
	u_{\omega}^{(\textsc{in})}(t)=\alpha_{\bm{k}}^{\vphantom{(\textsc{out})}}u_{\omega}^{(\textsc{out})}(t)+\beta_{\bm{k}}^{\vphantom{(\textsc{out})}}u_{\omega}^{(\textsc{out})*}(t)\,,
\end{equation}
after we factor out $e^{i\bm{k}\cdot\bm{x}}$. Thus, for an isotropic, spatially-flat space, the Bogoliubov coefficients depend only on $\lvert\bm{k}\rvert$ or $\omega$.

The annihilation operator $\hat{b}_{\bm{k}}$ associated with the OUT mode function $u_{\bm{k}}^{(\textsc{out})}$ will be related to the operators $(\hat{a}_{\bm{k}}^{\vphantom{\dagger}},\hat{a}_{\bm{k}}^{\dagger})$ by
\begin{equation}\label{E:ghjrbdfdf}
	\hat{b}_{\bm{k}}^{\vphantom{\dagger}}=\alpha_{\bm{k}}^{\vphantom{\dagger}}\,\hat{a}_{\bm{k}}^{\vphantom{\dagger}}+\beta_{-\bm{k}}^{\vphantom{\dagger}*}\,\hat{a}_{-\bm{k}}^{\dagger}\,.
\end{equation}
This implies that in terms of $\hat{b}_{\bm{k}}$ we have nonzero {\it particle number} $N_{\bm{k}}$ and {\it quantum coherence} $C_{\bm{k}}$
\begin{align}
	N_{\bm{k}}^{(\textsc{out})}&=\langle\hat{b}_{+\bm{k}}^{\dagger}\hat{b}_{+\bm{k}}^{\vphantom{\dagger}}\rangle=\lvert\beta_{\bm{k}}\rvert^{2}\,,\label{E:fghsif}\\
	C_{\bm{k}}^{(\textsc{out})}&=\langle\hat{b}_{+\bm{k}}\hat{b}_{-\bm{k}}\rangle=\alpha_{\bm{k}}^{\vphantom{*}}\beta_{\bm{k}}^{*}\,,\label{E:fghsif1}
\end{align}
in the vacuum state associated with $\hat{a}_{\bm{k}}$ if $\beta_{\bm{k}}\neq0$, while in comparison, in terms of $\hat{a}_{\bm{k}}$, they are
\begin{align}
	N_{\bm{k}}^{(\textsc{in})}&=\langle\hat{a}_{+\bm{k}}^{\dagger}\hat{a}_{+\bm{k}}^{\vphantom{\dagger}}\rangle=0\,,\\
	C_{\bm{k}}^{(\textsc{in})}&=\langle\hat{a}_{+\bm{k}}\hat{a}_{-\bm{k}}\rangle=0\,.
\end{align}
The physical meaning  of these results will be  explained  later.

We may choose $\alpha_{\bm{k}}$ in \eqref{E:rubses} to be real if we factor out the additional rotation in the transformation \eqref{E:rthdnf1}. This implies that generically the Bogoliubov coefficients can be parametrized by
\begin{align}\label{E:bgsgfdfsd}
	\alpha_{\bm{k}}&=\cosh\eta_{\bm{k}}\,,&\beta_{\bm{k}}&=-\sinh\eta_{\bm{k}}\,e^{-i\theta_{\bm{k}}}\,,
\end{align}
with the time- and mode-dependent parameters $\eta_{\bm{k}}>0$ and $0\leq\theta_{\bm{k}}<2\pi$. These expressions remind us the squeeze transformation of the creation and annihilation operators in quantum optics. Actually, we can write the linear transformation \eqref{E:ghjrbdfdf} as the two-mode squeezing of $\hat{a}_{\bm{k}}$ by the two-mode squeeze operator 
\begin{equation}
	\hat{S}_{2}^{\vphantom{\dagger}}(\zeta_{\bm{k}}^{\vphantom{\dagger}})=\exp\Bigl[\zeta_{\bm{k}}^{*\vphantom{\dagger}}\,\hat{a}_{\bm{k}}^{\vphantom{\dagger}}\hat{a}_{-\bm{k}}^{\vphantom{\dagger}}-\zeta_{\bm{k}}^{\vphantom{\dagger}}\,\hat{a}_{\bm{k}}^{\dagger}\hat{a}_{-\bm{k}}^{\dagger}\Bigr]\,,\label{E:bkeursd}
\end{equation}
with the squeeze parameter $\zeta_{\bm{k}}=\eta_{\bm{k}}\,e^{i\theta_{\bm{k}}}$ such that
\begin{align}\label{E:dgnskg}
	\hat{b}_{\bm{k}}^{\vphantom{\dagger}}&=\hat{S}_{2}^{\dagger}(\zeta_{\bm{k}}^{\vphantom{\dagger}})\,\hat{a}_{\bm{k}}^{\vphantom{\dagger}}\,\hat{S}_{2}^{\vphantom{\dagger}}(\zeta_{\bm{k}}^{\vphantom{\dagger}})=\cosh\eta_{\bm{k}}^{\vphantom{\dagger}}\,\hat{a}_{\bm{k}}^{\vphantom{\dagger}}-e^{+i\theta_{\bm{k}}}\sinh\eta_{\bm{k}}^{\vphantom{\dagger}}\,\hat{a}_{-\bm{k}}^{\dagger}\,.
\end{align}
When it applies to the vacuum state, it gives a two-mode squeezed vacuum state 
\begin{equation}
	\lvert\zeta_{\bm{k}}^{(2)}\rangle=\hat{S}_{2}^{\vphantom{\dagger}}(\zeta_{\bm{k}}^{\vphantom{\dagger}})\,\lvert0_{+\bm{k}},0_{-\bm{k}}\rangle=\frac{1}{\cosh\eta_{\bm{k}}}\sum_{n=0}^{\infty}\bigl(-\tanh\eta_{\bm{k}}\,e^{+i\theta_{\bm{k}}}\bigr)^{n}\lvert n_{+\bm{k}},n_{-\bm{k}}\rangle\,,\label{E:ngbserte}
\end{equation}
in terms of the in-particle number states, and the corresponding density matrix operator is given by
\begin{align}\label{E:roshejd}
	\hat{\rho}_{\bm{k}}=\lvert\zeta_{\bm{k}}^{(2)}\rangle\langle\zeta_{\bm{k}}^{(2)}\rvert\,,
\end{align}
still a pure state, but in general the off-diagonal terms $n\neq m$ of the density matrix do not vanish for each mode $\bm{k}$. Eq.~\eqref{E:ngbserte} explicitly shows that the resulting state contains particles in pairs, with opposite momenta. Note that Eq.~\eqref{E:bkeursd} implies that $\zeta_{\bm{k}}=\zeta_{-\bm{k}}$, so that we use $+\bm{k}$ instead of $-\bm{k}$ for the squeeze parameters in \eqref{E:bgsgfdfsd}. The identification \eqref{E:ngbserte} is compatible with the fact that the most general Gaussian state described by the density matrix operator \eqref{E:lrngflgs} is a squeezed thermal state.

Relating the Bogoliubov coefficients to the squeeze operator in \eqref{E:dgnskg} provides an active view of the Bogoliubov transformation. It allows a mapping of an operator in the Heisenberg picture at one time to its counterpart at another time, that is, under time evolution. Or, alternatively, in the Schr\"odinger picture it maps the time evolution of the state. This is in contrast to the implementation of the Bogoliubov coefficients in \eqref{E:rthdnf1}, where they are used to relate different bases (mode expansion) of the same operator. This is the passive view. Hereafter we will adopt the active view to investigate the time evolution of the parametric field.

Now from this active view, we see that the parametric process in general will drive the field from its initial vacuum state to a squeezed vacuum state, creating particles in pairs. It is also interesting to note that these $N_{\bm{k}}$ and $C_{\bm{k}}$ account for three parameters, needed to fully specify a Gaussian state, as $\mathsf{a}_{\bm{k}}$, $\mathsf{b}_{\bm{k}}$, and $\mathsf{c}_{\bm{k}}$ do, and thus they contain complete information of the linear quantum parametric field in a Gaussian state via the numbers of particle it creates and the correlation between the produced  particle pairs.

The particles, as seen by $\hat{b}_{\bm{k}}$, are correlated. According to the active view, Eq.~\eqref{E:fghsif} only accounts for spontaneous creation since we have assumed that the initial state is a vacuum. If the initial state is not a vacuum state,  but with $\langle\hat{N}_{\bm{k}}^{(\textsc{in})}\rangle$ particles present we can show that
\begin{align}
	\sum_{\bm{k}>0}\langle\hat{N}_{+\bm{k}}^{(\textsc{out})}\rangle+\langle\hat{N}_{-\bm{k}}^{(\textsc{out})}\rangle=\sum_{\bm{k}}2\Bigl(\lvert\beta_{\bm{k}}^{\vphantom{\dagger}}\rvert^{2}+\frac{1}{2}\Bigr)\Bigl(\langle\hat{N}_{\bm{k}}^{(\textsc{in})}\rangle+\frac{1}{2}\Bigr)-\frac{1}{2}\,,
\end{align}
where $\langle\cdots\rangle$ are taken with respect to the initial state.  Thus in addition to spontaneous particle creation, there is also  stimulated creation due to the presence of particles in the initial state.

So far we have given a  discussion of the generic kinematical properties of particle creation in a time-dependent background and made identification with quantum squeezing by the language of  Bogoliubov transformations. We now   show  their connection with dynamics by using a nonequilibrium quantum dynamics  formulation.

\subsection{Nonequilibrium evolution of modes in time }

In the Heisenberg picture, the time evolution of the parametric field is accounted for by the (field) operators, and for the linear field, this information is encoded in the mode function $u_{\omega}(t)$. In other words, the time evolution of an operator can be translated into the time evolution of the mode function that span the operator. As an initial-value problem, in general, the mode function $u_{\omega}(t)$ at any time can be expanded by its initial conditions at $t=0$ by
\begin{equation}\label{E:dkhedfd}
	u_{\omega}(t)=d_{\omega}^{(1)}(t)\,u_{\omega}(0)+d_{\omega}^{(2)}(t)\,\dot{u}_{\omega}(0)\,,
\end{equation}
where the overdot represents the time derivative. The two functions $d_{\omega}^{(1)}(t)$ and $d_{\omega}^{(2)}(t)$ are a special set of homogeneous solutions to the wave equation \eqref{E:fgksgf}, satisfying
\begin{align}
	d_{\omega}^{(1)}(0)&=1\,,&\dot{d}_{\omega}^{(1)}(0)&=0\,,&d_{\omega}^{(2)}(0)&=0\,,&\dot{d}_{\omega}^{(2)}(0)&=1\,.
\end{align}
This approach is convenient in the sense that for a linear system, almost all physical observables can be packaged into these two functions plus some initial conditions, so it would be computationally efficient once we get hold of the functions $d_{\omega}^{(1)}(t)$ and $d_{\omega}^{(2)}(t)$.

For convenience, suppose that the in-mode functions have the form 
\begin{align}\label{E:rtrhitr}
	u_{\omega}^{(\textsc{in})}(t)&=\frac{1}{\sqrt{2\omega_{i}}}\,e^{-i\omega_{i}t}\,,&&\text{and}&\dot{u}_{\omega}^{(\textsc{in})}(t)&=-i\sqrt{\frac{\omega_{i}}{2}}\,e^{-i\omega_{i}t}\,,
\end{align}
In terms for the fundamental solutions $d_{\omega}^{(i)}$, we can write the out-mode functions as
\begin{equation}\label{E:etsfsdf}
	u_{\omega}^{(\textsc{out})}(t)=\frac{1}{\sqrt{2\omega_{i}}}\,\bigl[d_{\omega}^{(1)}(t)-i\omega_{i}d_{\omega}^{(2)}(t)\bigr]\,,
\end{equation}
according to \eqref{E:dkhedfd}, where $\omega_{i}=\omega(0)$ is the initial value of the frequency modulation. Thus the field operator $\hat{\phi}(\bm{x},t)$ at time $t$ will be given by
\begin{align}
	\hat{\phi}(\bm{x},t)=\int\!\frac{d^{3}k}{(2\pi)^{\frac{3}{2}}}\frac{1}{\sqrt{2\omega_{i}}}\Bigl\{\hat{a}_{\bm{k}}^{\vphantom{\dagger}}\,e^{+i\bm{k}\cdot\bm{x}}\bigl[d_{\omega}^{(1)}(t)-i\omega_{i}\,d_{\omega}^{(2)}(t)\bigr]+\hat{a}_{\bm{k}}^{\dagger}\,e^{-i\bm{k}\cdot\bm{x}}\bigl[d_{\omega}^{(1)}(t)+i\omega_{i}\,d_{\omega}^{(2)}(t)\bigr]\Bigr\}\,,\label{E:bgdker}
\end{align}
in contrast to the in-field before the parametric process starts at $t=0$ 
\begin{align}
	\hat{\phi}_{\textsc{in}}(\bm{x},t)=\int\!\frac{d^{3}k}{(2\pi)^{\frac{3}{2}}}\frac{1}{\sqrt{2\omega_{i}}}\Bigl\{\hat{a}_{\bm{k}}^{\vphantom{\dagger}}\,e^{+i\bm{k}\cdot\bm{x}}\,e^{-i\omega_{i}t}+\hat{a}_{\bm{k}}^{\dagger}\,e^{-i\bm{k}\cdot\bm{x}}\,e^{+i\omega_{i}t}\Bigr\}\,.
\end{align}
When we apply the two-mode squeezing to this in-field, we can relate them by
\begin{align}\label{E:rohdfd}
	\hat{\phi}(\bm{x},t)&=\hat{S}_{2}^{\dagger}(\{\zeta_{\bm{k}}\})\,\hat{\phi}_{\textsc{in}}(\bm{x},t)\,\hat{S}_{2}(\{\zeta_{\bm{k}}\})\\
	&=\int\!\!\frac{d^{3}\bm{k}}{(2\pi)^{\frac{3}{2}}}\;\frac{1}{\sqrt{2\omega_{i}}}\,\Bigl\{\Bigl[\alpha_{\bm{k}}^{\vphantom{*}}\,e^{-i\omega_{i}t}+\beta_{\bm{k}}^{\vphantom{*}}\,e^{+i\omega_{i}t}\Bigr]\,\hat{a}^{\vphantom{\dagger}}_{\bm{k}}\,e^{+i\bm{k}\cdot\bm{x}}+\Bigl[\beta_{\bm{k}}^{*}\,e^{-i\omega_{i}t}+\alpha_{\bm{k}}^{*}\,e^{+i\omega_{i}t}\Bigr]\,\hat{a}^{\dagger}_{\bm{k}}\,e^{-i\bm{k}\cdot\bm{x}}\Bigr\}\,,\notag
\end{align}
where $\hat{S}_{2}(\{\zeta_{\bm{k}}\})$ is understood as a collection of two-mode squeeze operators for all $\bm{k}>0$. Compare this expression with \eqref{E:bgdker} and we find
\begin{align}\label{E:fnljsdb1}
	u_{\omega}^{(\textsc{out})}(t)&=\alpha_{\bm{k}}^{\vphantom{*}}(t)\,u_{\omega}^{(\textsc{in})}(t)+\beta_{\bm{k}}^{\vphantom{*}}(t)\,u_{\omega}^{(\textsc{in})*}(t)\,,\notag\\
	\Rightarrow\qquad\qquad d^{(1)}_{\omega}(t)-i\,\omega_{i}\,d^{(2)}_{\omega}(t)&=e^{-i\omega_{i}t}\,\alpha_{\bm{k}}^{\vphantom{*}}(t)+e^{+i\omega_{i}t}\,\beta_{\bm{k}}^{\vphantom{*}}(t)\,,
\end{align}
for $t>0$. Applying similar arguments to the conjugate momentum $\hat{\pi}(\bm{x},t)$, we will obtain
\begin{align}\label{E:fnljsdb2}
	\dot{u}_{\omega}^{(\textsc{out})}(t)&=\alpha_{\bm{k}}^{\vphantom{*}}(t)\,\dot{u}_{\omega}^{(\textsc{in})}(t)+\beta_{\bm{k}}^{\vphantom{*}}(t)\,\dot{u}_{\omega}^{(\textsc{in})*}(t)\,,\notag\\
	\Rightarrow\qquad\qquad\dot{d}^{(1)}_{\omega}(t)-i\,\omega_{i}\,\dot{d}^{(2)}_{\omega}(t)&=-i\,\omega_{i}\,e^{-i\omega_{i}t}\,\alpha_{\bm{k}}(t)+i\,\omega_{i}\,e^{+i\omega_{i}t}\,\beta_{\bm{k}}(t)\,.
\end{align}
Solving \eqref{E:fnljsdb1} and \eqref{E:fnljsdb2} simultaneously we arrive at
\begin{align}
	\alpha_{\bm{k}}(t)&=\frac{1}{2\omega_{i}}\,e^{+i\omega_{i}t}\,\Bigl[\omega_{i}\,d^{(1)}_{\omega}(t)+i\,\dot{d}^{(1)}_{\omega}(t)-i\,\omega_{i}^{2}d^{(2)}_{\omega}(t)+\omega_{i}\,\dot{d}^{(2)}_{\omega}(t)\Bigr]\,,\label{E:gbkrjbkdg1}\\
	\beta_{\bm{k}}(t)&=\frac{1}{2\omega_{i}}\,e^{-i\omega_{i}t}\,\Bigl[\omega_{i}\,d^{(1)}_{\omega}(t)-i\,\dot{d}^{(1)}_{\omega}(t)-i\,\omega_{i}^{2}d^{(2)}_{\omega}(t)-\omega_{i}\,\dot{d}^{(2)}_{\omega}(t)\Bigr]\,.\label{E:gbkrjbkdg2}
\end{align}
Thus we have expressed the Bogoliubov coefficients of all modes at any time by the corresponding fundamental solutions $d^{(i)}_{\omega}(t)$. Then from \eqref{E:bgsgfdfsd}, we can write the squeeze parameters by the same set of fundamental solutions. For example, we can show
\begin{equation}
	\lvert\alpha_{\bm{k}}\rvert^{2}+\lvert\beta_{\bm{k}}\rvert^{2}=\frac{1}{2\omega_{i}^{2}}\Big[\omega_{i}^{2}d^{(1)2}_{\omega}(t)+\dot{d}^{(1)2}_{\omega}(t)+\omega_{i}^{4}d^{(2)2}_{\omega}(t)+\omega_{i}^{2}\dot{d}^{(2)2}_{\omega}(t)\Bigr]=\cosh2\eta_{\bm{k}}\,.
\end{equation}
This also implies that the squeeze parameters $\eta_{\bm{k}}$, $\theta_{\bm{k}}$ are actually functions of $\lvert\bm{k}\rvert$.

The covariance matrix elements $\mathsf{b}_{\bm{k}}$, $\mathsf{a}_{\bm{k}}$ and $\mathsf{c}_{\bm{k}}$ can also be expressed by these fundamental solutions via the Bogoliubov coefficients. For example, from \eqref{E:rkjgb1} and \eqref{E:rbgrivfghs}, we have
\begin{align}
	\hat{\phi}_{\bm{k}}^{(1)}(t)=\frac{1}{\sqrt{2}}\Bigl[\hat{a}_{+\bm{k}}^{\vphantom{\dagger}}u_{\omega}^{\vphantom{*}(\textsc{out})}(t)+\hat{a}_{-\bm{k}}^{\dagger}u_{\omega}^{(\textsc{out})*}(t)+\hat{a}_{-\bm{k}}^{\vphantom{\dagger}}u_{\omega}^{(\textsc{out})\vphantom{*}}(t)+\hat{a}_{+\bm{k}}^{\dagger}u_{\omega}^{(\textsc{out})*}(t)\Bigr]\,,
\end{align}
and thus at any time
\begin{align}
	\mathsf{b}_{\bm{k}}&=\langle\hat{\phi}_{\bm{k}}^{(1)2}\rangle=u_{\omega}^{(\textsc{out})\vphantom{*}}u_{\omega}^{(\textsc{out})*}=2\Bigl\{\lvert\beta_{\bm{k}}\rvert^{2}+\frac{1}{2}\Bigr\}\,u_{\omega}^{(\textsc{in})}u_{\omega}^{(\textsc{in})*}+\alpha_{\bm{k}}^{\vphantom{*}}\beta_{\bm{k}}^{*}\,u_{\omega}^{(\textsc{in})2}+\alpha_{\bm{k}}^{*}\beta_{\bm{k}}^{\vphantom{*}}\,u_{\omega}^{(\textsc{in})*2}\,,\label{E:gkhjbsir1}\\
	\mathsf{a}_{\bm{k}}&=\langle p_{\bm{k}}^{(1)2}\rangle=\dot{u}_{\omega}^{(\textsc{out})\vphantom{*}}\dot{u}_{\omega}^{(\textsc{out})*}=2\Bigl\{\lvert\beta_{\bm{k}}\rvert^{2}+\frac{1}{2}\Bigr\}\,\dot{u}_{\omega}^{(\textsc{in})}\dot{u}_{\omega}^{(\textsc{in})*}+\alpha_{\bm{k}}^{\vphantom{*}}\beta_{\bm{k}}^{*}\,\dot{u}_{\omega}^{(\textsc{in})2}+\alpha_{\bm{k}}^{*}\beta_{\bm{k}}^{\vphantom{*}}\,\dot{u}_{\omega}^{(\textsc{in})*2}\,,\\
	\mathsf{c}_{\bm{k}}&=\frac{1}{2}\langle\bigl\{\phi_{\bm{k}}^{(1)},p_{\bm{k}}^{(1)}\bigr\}\rangle=\frac{1}{2}\biggl\{\dot{u}_{\omega}^{(\textsc{out})\vphantom{*}}u_{\omega}^{(\textsc{out})*}+u_{\omega}^{(\textsc{out})\vphantom{*}}\dot{u}_{\omega}^{(\textsc{out})*}\biggr\}\label{E:gkhjbsir3}\\
	&\qquad\qquad\qquad\quad\;\;\,=\Bigl\{\lvert\beta_{\bm{k}}\rvert^{2}+\frac{1}{2}\Bigr\}\,\frac{d}{dt}\Bigl[u_{\omega}^{(\textsc{in})}u_{\omega}^{(\textsc{in})*}\Bigr]+\frac{1}{2}\,\alpha_{\bm{k}}^{\vphantom{*}}\beta_{\bm{k}}^{*}\,\frac{d}{dt}u_{\omega}^{(\textsc{in})2}+\frac{1}{2}\,\alpha_{\bm{k}}^{*}\beta_{\bm{k}}^{\vphantom{*}}\,\frac{d}{dt}u_{\omega}^{(\textsc{in})*2}\,,\notag
\end{align}
Then we can further use \eqref{E:gbkrjbkdg1} and \eqref{E:gbkrjbkdg2} or simply use only \eqref{E:etsfsdf} and \eqref{E:fnljsdb2}. If the in-mode function is of the form \eqref{E:rtrhitr}, we can easily relate the covariance matrix elements with the particle number and coherence \eqref{E:fghsif} and \eqref{E:fghsif1} in the out-state by
\begin{align}
	\mathsf{b}_{\bm{k}}(t)&=d_{\omega}^{(1)2}(t)\,\frac{1}{2\omega_{i}}+d_{\omega}^{(2)2}(t)\,\frac{\omega_{i}}{2}\\
	&=\bigl(N_{\bm{k}}^{\textsc{(out)}}(t)+\frac{1}{2}\bigr)\,\frac{1}{\omega_{i}}+\frac{1}{2\omega_{i}}\,C_{\bm{k}}^{\textsc{(out)}\vphantom{*}}(t)\,e^{-i2\omega_{i}t}+\frac{1}{2\omega_{i}}\,C_{\bm{k}}^{\textsc{(out)}*}(t)\,e^{+i2\omega_{i}t}\,,\\
	\mathsf{a}_{\bm{k}}(t)&=\dot{d}_{\omega}^{(1)2}(t)\,\frac{1}{2\omega_{i}}+\dot{d}_{\omega}^{(2)2}(t)\,\frac{\omega_{i}}{2}\\
	&=\bigl(N_{\bm{k}}^{\textsc{(out)}}(t)+\frac{1}{2}\bigr)\,\omega_{i}-\frac{\omega_{i}}{2}\,C_{\bm{k}}^{\textsc{(out)}\vphantom{*}}(t)\,e^{-i2\omega_{i}t}-\frac{\omega_{i}}{2}\,C_{\bm{k}}^{\textsc{(out)}*}(t)\,e^{+i2\omega_{i}t}\,,\\
	\mathsf{c}_{\bm{k}}(t)&=d_{\bm{k}}^{(1)}(t)\dot{d}_{\omega}^{(1)}(t)\,\frac{1}{2\omega_{i}}+d_{\omega}^{(2)}(t)\dot{d}_{\bm{k}}^{(2)}(t)\,\frac{\omega_{i}}{2}\\
	&=-\frac{i}{2}\,C_{\bm{k}}^{\textsc{(out)}\vphantom{*}}(t)\,e^{-i2\omega_{i}t}+\frac{i}{2}\,C_{\bm{k}}^{\textsc{(out)}*}(t)\,e^{+i2\omega_{i}t}\,,
\end{align}
or vice versa, write the particle number and coherence in terms of the fundamental solutions, attaining  {in principle} all detailed information of their time evolution
\begin{figure}
	\centering
    \scalebox{0.5}{\includegraphics{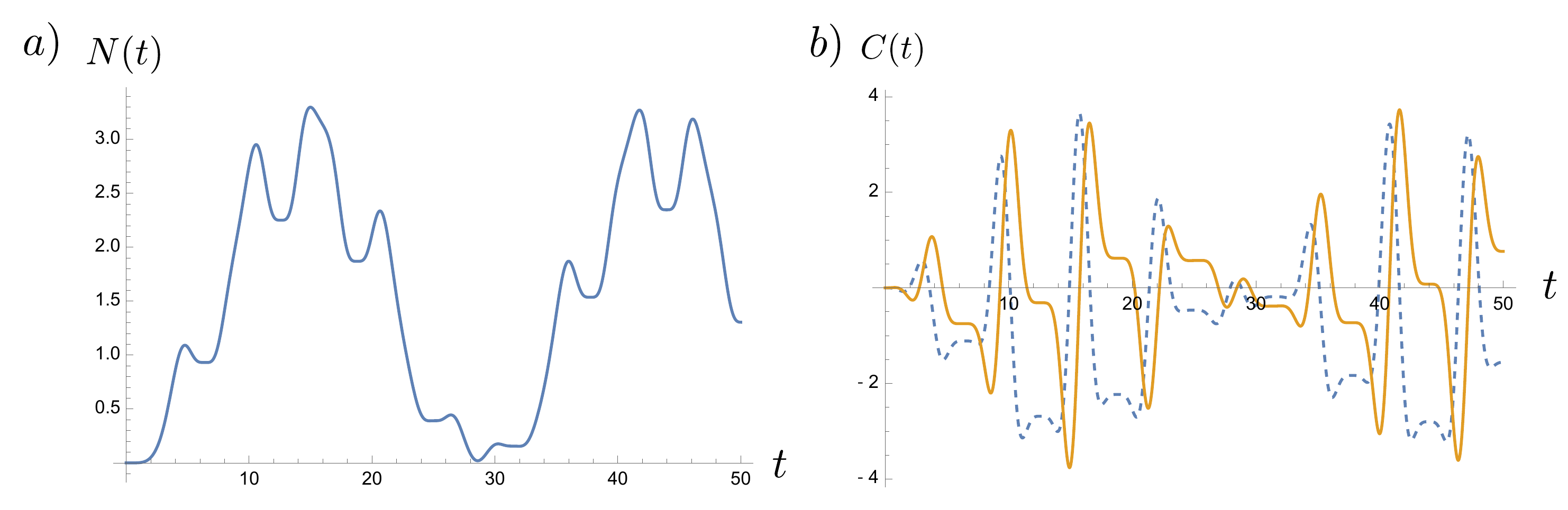}}
    \caption{The time variation of the particle number $N(t)$ and coherence $C(t)$ for the parametric oscillator having the frequency modulation given by \eqref{E:dfbksre}. We choose $\mathfrak{R}=1$ and $\varpi=0.5$. In (b) the blue dashed curve represents the real part of $C(t)$ and the orange solid curve denotes the imaginary part.}\label{Fi:NC1}
\end{figure}
\begin{align}
	N_{\bm{k}}^{(\textsc{out})}(t)&=\sinh^{2}\eta_{\bm{k}}(t)=\frac{1}{4}\Bigl\{\Bigl[d_{\omega}^{(1)}(t)-\dot{d}_{\omega}^{(2)}(t)\Bigr]^{2}+\Bigl[\omega_{i}d_{\omega}^{(2)}(t)+\frac{1}{\omega_{i}}\,\dot{d}_{\omega}^{(1)}(t)\Bigr]^{2}\Bigr\}\notag\\
	&=\frac{1}{4}\Bigl[d_{\omega}^{(1)2}(t)+\frac{1}{\omega_{i}^{2}}\,\dot{d}_{\omega}^{(1)2}(t)+\omega_{i}^{2}d_{\omega}^{(2)2}(t)+\dot{d}_{\omega}^{(2)2}(t)\Bigr]-\frac{1}{2}\,,\label{E:fkgjbdjf1}\\
\intertext{and}
	C_{\bm{k}}^{(\textsc{out})}(t)&=-\frac{1}{2}\,\sinh2\eta_{\bm{k}}(t)\,e^{+i\theta_{\bm{k}}(t)}=\frac{e^{i2\omega_{i}t}}{4}\Bigl\{\Bigl[d_{\bm{k}}^{(1)}(t)+\frac{i}{\omega_{i}}\,\dot{d}_{\bm{k}}^{(1)}(t)\Bigr]^{2}+\omega_{i}^{2}\Bigl[d_{\bm{k}}^{(2)}(t)+\frac{i}{\omega_{i}}\,\dot{d}_{\bm{k}}^{(2)}(t)\Bigr]^{2}\Bigr\}\,.\label{E:fkgjbdjf2}
\end{align}
We observe that comparing with $N_{\bm{k}}^{(\textsc{out})}(t)$, the coherence $C_{\bm{k}}^{(\textsc{out})}(t)$ tends to oscillate more rapidly with time (but not necessarily always so),  even though in principle both particle number and coherence are oscillatory, because $d_{\omega}^{(i)}$ are solutions of the parametric oscillator.  {To explicitly illustrate this point, we  give a few examples. First consider a parametric oscillator with the time-dependent frequency 
\begin{equation}\label{E:dfbksre}
	\omega^{2}(t)=\mathfrak{R}^{2} \cos^{2}\varpi t\,,
\end{equation}
where $\varpi$ is the frequency of  modulation and $\mathfrak{R}$ is the amplitude. From Fig.~\ref{Fi:NC1}, we find that both $N(t)$ and $C(t)$ are oscillatory more or less at the same tempo except that $N(t)$ is always nonnegative. A  contrasting example is the case with frequency 
\begin{equation}\label{E:oertkfdj}
	\omega^{2}(t)=\mathfrak{R}^{2} \bigl(1-e^{-2\varpi t}\bigr)\,,
\end{equation}
where the frequency modulation monotonically rises from zero to a constant $\mathfrak{R}$. In this case both $N(t)$ and $C(t)$ barely oscillate and quickly settle down to constant values. These examples illustrate our point that   the contributions of quantum coherence need be taken more seriously, and arguments to justify its cancellation due to rapid oscillations of coherence in comparison to particle number may be too simplistic.
\begin{figure}
	\centering
    \scalebox{0.5}{\includegraphics{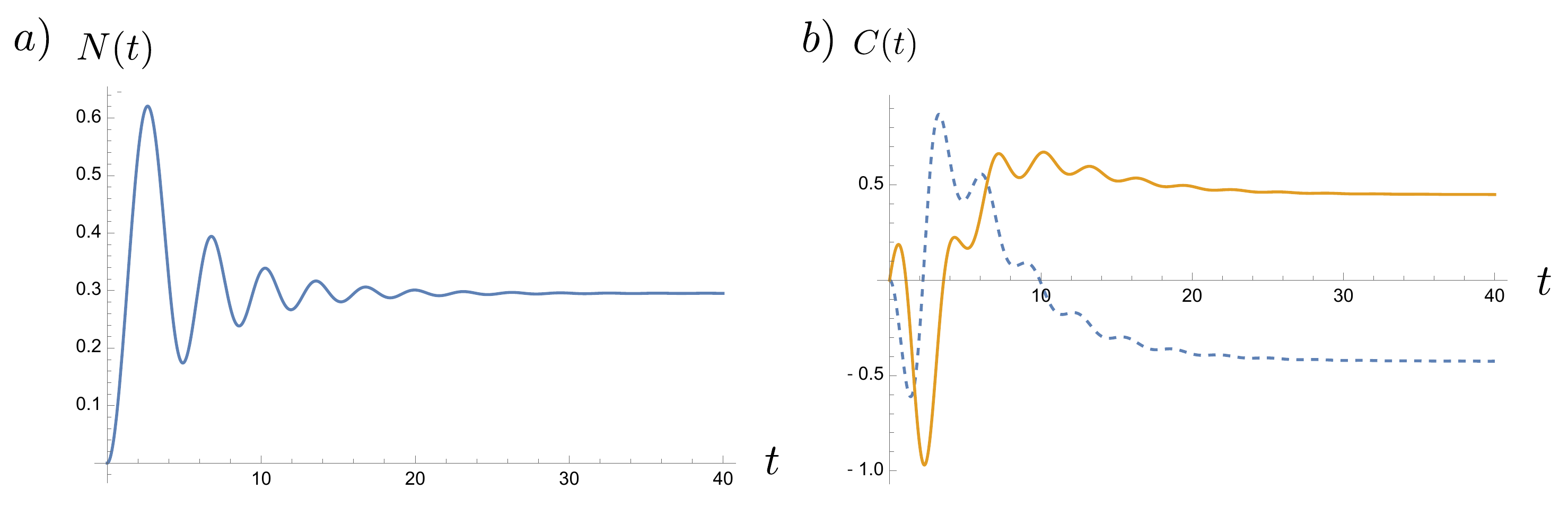}}
    \caption{The time variation of the particle number $N(t)$ and coherence $C(t)$ for the parametric oscillator having the frequency modulation given by \eqref{E:oertkfdj}. We choose $\mathfrak{R}=1$ and $\varpi=0.1$. In (b) the blue dashed curve represents the real part of $C(t)$ and the orange solid curve denotes the imaginary part.}\label{Fi:NC2}
\end{figure}
}

For a Gaussian system, the physical quantities we are interested in can be built from the covariance matrix elements. We shall now focus on the von Neumann entropy.

\section{von Neumann entropy of the parametric quantum field}\label{S:roithnd}

The von Neumann entropy $\mathcal{S}_{\bm{k}}$ for $(\phi_{\bm{k}}^{(1)},p_{\bm{k}}^{(1)})$ 
\begin{equation}\label{E:kfhjdfd}
	\mathcal{S}_{\bm{k}}=-\operatorname{Tr}\Bigl\{\hat{\rho}(\hat{\phi}_{\bm{k}}^{(1)},\hat{\pi}_{\bm{k}}^{(1)})\,\ln\hat{\rho}(\hat{\phi}_{\bm{k}}^{(1)},\hat{\pi}_{\bm{k}}^{(1)})\Bigr\}
\end{equation}
can be given by the symplectic eigenvalue\footnote{As a reminder, the symplectic eigenvalue $\lambda$ of a $2N$-dimensional covariance matrix $\bm{\sigma}$ can be found by solving the eigenvalue problem of the form
\begin{equation*}
	\bm{\sigma}\cdot\bm{v}=\lambda\,\bm{\Sigma}\cdot\bm{v}\,,\qquad\text{or equivalently}\qquad \bigl(\bm{\Sigma}\cdot\bm{\sigma}\bigr)\cdot\bm{v}=\lambda\,\bm{v}\,,
\end{equation*}
where the matrix $\bm{\Sigma}$ is the fundamental symplectic matrix
\begin{equation*}
	\bm{\Sigma}=\bigoplus_{k=1}^{N}\begin{pmatrix} 0 &i\\-i &0\end{pmatrix}\,.
\end{equation*}} of the covariance matrix $\bm{\sigma}$ associated with the canonical variables $(\phi_{\bm{k}}^{(1)},p_{\bm{k}}^{(1)})$
\begin{equation}\label{E:fgjhdfgdx}
	\mathcal{S}_{\bm{k}}=\bigl(\lambda_{\bm{k}}+\frac{1}{2}\bigr)\,\ln\bigl(\lambda_{\bm{k}}+\frac{1}{2}\bigr)-\bigl(\lambda_{\bm{k}}-\frac{1}{2}\bigr)\,\ln\bigl(\lambda_{\bm{k}}-\frac{1}{2}\bigr)\,,
\end{equation}
where the symplectic eigenvalue $\lambda_{\bm{k}}$ is defined by
\begin{equation}\label{E:kjfger}
	\lambda_{\bm{k}}^{2}=\mathsf{a}_{\bm{k}}^{\vphantom{*}}\mathsf{b}_{\bm{k}}^{\vphantom{*}}-\mathsf{c}_{\bm{k}}^{2}=\langle\phi_{\bm{k}}^{(1)2}\rangle\langle p_{\bm{k}}^{(1)2}\rangle-\Bigl[\frac{1}{2}\langle\bigl\{\phi_{\bm{k}}^{(1)},p_{\bm{k}}^{(1)}\bigr\}\rangle\Bigr]{}^{2}\,.
\end{equation}
From the covariance matrix elements \eqref{E:gkhjbsir1}--\eqref{E:gkhjbsir3}, if the in-mode functions satisfy the Wronskian conditions, then we find the symplectic eigenvalues can be expressed by $N_{\bm{k}}$ and $C_{\bm{k}}$ as well
\begin{align}
	\lambda_{\bm{k}}^{2}=\mathsf{a}_{\bm{k}}^{\vphantom{*}}\mathsf{b}_{\bm{k}}^{\vphantom{*}}-\mathsf{c}_{\bm{k}}^{2}&=-\Bigl[\bigl(N_{\bm{k}}^{(\textsc{out})}+\frac{1}{2}\bigr)^{2}-\lvert C_{\bm{k}}^{(\textsc{out})}\rvert^{2}\Bigr]\Bigl[u_{k}^{(\textsc{in})}\dot{u}_{k}^{(\textsc{in})*}-\dot{u}_{k}^{(\textsc{in})}u_{k}^{(\textsc{in})*}\Bigr]{}^{2}\notag\\
	&=\bigl(N_{\bm{k}}^{(\textsc{out})}+\frac{1}{2}\bigr)^{2}-\lvert C_{\bm{k}}^{(\textsc{out})}\rvert^{2}\,,\label{E:fgkjdkfg}
\end{align}
independent of the phase of $C_{\bm{k}}^{(\textsc{out})}$ and independent of the explicit expressions of the in-mode functions. The latter merely reflects the invariance of the von Neumann entropy of the system during its unitary time evolution.  {The former implies that the phase of the coherence plays no role in evaluating the symplectic eigenvalue, and thus the von Neumann entropy. It further suggests we should not assume that a rapidly oscillating coherence can be canceled out and smear off its contribution to the von Neumann entropy.}

Now we see that the entropy thus generated is formally somewhat different from that of a thermal state of the harmonic oscillator, or of a photon mode, we often come across
\begin{align}\label{E:kjngskd}
	\mathcal{S}_{\bm{k}}&=\bigl(N_{\bm{k}}+1\bigr)\,\ln\bigl(N_{\bm{k}}+1\bigr)-N_{\bm{k}}\,\ln N_{\bm{k}}\,,&N_{\bm{k}}&=\frac{1}{e^{\beta\omega_{i}}-1}\,,
\end{align}
by an additional coherence $\lvert C_{\bm{k}}\rvert^{2}$. The expressions of $N_{\bm{k}}$ is different too. At first sight it is rather baffling why there exists any entropy \eqref{E:fgjhdfgdx} for a parametric process from a vacuum state, since the final state of each mode is a pure two-mode squeezed state. It is irrelevant to the basis we use in calculating the trace in \eqref{E:kfhjdfd} because the entropy is defined in terms of the trace and the eigenvalues $\lambda_{\bm{k}}$ are invariant under the linear symplectic transformations. It is of importance to reconcile the paradoxical coexistence of the time-reversal of unitary evolution and the time irreversible entropy production.

The entropy \eqref{E:fgjhdfgdx}, together with \eqref{E:kjfger}, seems to have a nonzero value, but the squeeze transformation alone does not change the value of the uncertainty function, that is, $\lambda_{\bm{k}}^{2}$, so if the initial state before the parametric process is a vacuum state, whose entropy vanishes, then the resulting squeezed vacuum state due to the parametric process will also have zero entropy. This is consistent with the fact that both the vacuum state and the squeezed vacuum state are pure states, and thus their entropies are zero. Hence we expect \eqref{E:fgjhdfgdx} should vanish. Alternatively, since the density matrix of a closed system is governed by the Liouville equation
\begin{equation}
	\frac{\partial\hat{\rho}(t)}{\partial t}=-i\,\bigl[\hat{H}(t),\hat{\rho}(t)\bigr]\,,
\end{equation}
we can explicitly show that the von Neumann entropy is a constant of motion
\begin{align}
	\frac{d}{dt}\mathcal{S}(t)=-\frac{d}{dt}\operatorname{Tr}\hat{\rho}(t)\ln\hat{\rho}(t)=i\,\operatorname{Tr}\Bigl\{\bigl[\hat{H}(t),\hat{\rho}(t)\bigr]\,\ln\hat{\rho}(t)\Bigr\}=0\,,
\end{align}
where we have used the cyclic property of the trace and
\begin{align}
	\operatorname{Tr}\hat{\rho}(t)&=1\,,&&\Rightarrow&\operatorname{Tr}\frac{\partial\hat{\rho}(t)}{\partial t}&=0\,.
\end{align}
That is, the entropy is indeed conserved under the unitary evolution. It implies that even the number of particles is changing, the entropy $\mathcal{S}$ remains the same.

On the other hand from \eqref{E:fkgjbdjf1} and \eqref{E:fkgjbdjf2} we can explicitlt verify that indeed \eqref{E:fgkjdkfg} is
\begin{equation}
	\lambda_{\bm{k}}^{2}=\bigl(N_{\bm{k}}^{(\textsc{out})}+\frac{1}{2}\bigr)^{2}-\lvert C_{\bm{k}}^{(\textsc{out})}\rvert^{2}=\frac{1}{4}\Bigl[d_{\bm{k}}^{(1)}(t)\dot{d}_{\bm{k}}^{(2)}(t)-\dot{d}_{\bm{k}}^{(1)}(t)d_{\bm{k}}^{(2)}(t)\Bigr]^{2}=\frac{1}{4}\,,\label{E:ioigfie}
\end{equation}
so that the corresponding entropy \eqref{E:fgjhdfgdx} of each field mode due to the parametric process does give zero, consistent with the earlier arguments. Thus we definitely will not have any entropy production for the parametric process evolving from the initial vacuum state.


At this point we have assumed that the field evolve unitarily, so the field evolves from a pure state into another one. We also assume that we can have full access to every bit  of  information   the field possesses.  {However, in reality, a physical system is never perfectly closed, but coupled to some environment.   When some  information of the system  is lost to its environment, or if some coarse-graining measure is introduced in the measurement protocols,  entropy is observed to be generated in the reduced system. When a  system undergoes nonequilibrium evolution some information is likely to be lost to the environment possessing a much larger number of degrees of freedom. The key point is coarse-graining,  resulting in certain information becoming irretrievable.  If we coarse-grain the information in the field we shall see nonzero entropy generated in the evolution of a pure state. }

\section{Evolution of the Wigner Function of the Quantum Field}
\label{S:wigner}
In the context of  creation of particle pairs, the Wigner function of the field offers the clearest picture of how the coarse-grained entropy may emerge, and a phase-space description offers an intimate view of its connection with quantum thermodynamics. 

\subsection{Wigner function of the field mode}
The Wigner function can be related to the density matrix elements \eqref{E:gdxkjs} by a Fourier transformation, 
\begin{equation}\label{E:rtsw}
	\mathcal{W}_{\bm{k}}^{(i)}(\Sigma_{\bm{k}}^{(i)},P_{\bm{k}}^{(i)};t)=\int\frac{du}{2\pi}\;e^{-iP_{\bm{k}}^{(i)}\Delta_{\bm{k}}^{(i)}}\,\rho(\Sigma_{\bm{k}}^{(i)}+\frac{\Delta_{\bm{k}}^{(i)}}{2},\Sigma_{\bm{k}}^{(i)}-\frac{\Delta_{\bm{k}}^{(i)}}{2};t)\,,
\end{equation}
and we obtain
\begin{equation}\label{E:kfghbs}
	\mathcal{W}_{\bm{k}}^{(i)}(\Sigma_{\bm{k}}^{(i)},P_{\bm{k}}^{(i)};t)=\frac{1}{2\pi\sqrt{\mathsf{a}_{\bm{k}}^{\vphantom{2}}\mathsf{b}_{\bm{k}}^{\vphantom{2}}-\mathsf{c}_{\bm{k}}^{2}}}\,\exp\Bigl[-\frac{\mathsf{b}_{\bm{k}}^{\vphantom{2}}P_{\bm{k}}^{(i)2}-2\mathsf{c}_{\bm{k}}^{\vphantom{2}}P_{\bm{k}}^{(i)}\Sigma_{\bm{k}}^{(i)}+\mathsf{a}_{\bm{k}}^{\vphantom{2}}\Sigma_{\bm{k}}^{(i)2}}{2(\mathsf{a}_{\bm{k}}^{\vphantom{2}}\mathsf{b}_{\bm{k}}^{\vphantom{2}}-\mathsf{c}_{\bm{k}}^{2})}\Bigr]\,.
\end{equation}
The parameters $N_{\bm{k}}$ and $C_{\bm{k}}$ include the information we need about the Gaussian state of the linear parametric field. If we make the substitutions
\begin{align}
	\mathsf{b}_{\bm{k}}&=2\bigl(N_{\bm{k}}^{\textsc{(out)}}+\frac{1}{2}\bigr)\,u_{\omega}^{(\textsc{in})}u_{\omega}^{(\textsc{in})*}+C_{\bm{k}}^{\textsc{(out)}\vphantom{*}}\,u_{\omega}^{(\textsc{in})2}+C_{\bm{k}}^{\textsc{(out)}*}\,u_{\omega}^{(\textsc{in})*2}\,,\label{E:fgnkde1}\\
	\mathsf{a}_{\bm{k}}&=2\bigl(N_{\bm{k}}^{\textsc{(out)}}+\frac{1}{2}\bigr)\,\dot{u}_{\omega}^{(\textsc{in})}\dot{u}_{\omega}^{(\textsc{in})*}+C_{\bm{k}}^{\textsc{(out)}\vphantom{*}}\,\dot{u}_{\omega}^{(\textsc{in})2}+C_{\bm{k}}^{\textsc{(out)}*}\,\dot{u}_{\omega}^{(\textsc{in})*2}\,,\\
	\mathsf{c}_{\bm{k}}&=\bigl(N_{\bm{k}}^{\textsc{(out)}}+\frac{1}{2}\bigr)\bigl(\dot{u}_{\omega}^{(\textsc{in})}u_{\omega}^{(\textsc{in})*}+u_{\omega}^{(\textsc{in})}\dot{u}_{\omega}^{(\textsc{in})*}\bigr)+C_{\bm{k}}^{\textsc{(out)}\vphantom{*}}\,u_{\omega}^{(\textsc{in})}\dot{u}_{\omega}^{(\textsc{in})}+C_{\bm{k}}^{\textsc{(out)}*}\,u_{\omega}^{(\textsc{in})*}\dot{u}_{\omega}^{(\textsc{in})*}\,,\label{E:fgnkde3}
\end{align}
and use the facts
\begin{align}
	u_{\omega}^{(\textsc{in})\vphantom{*}}(t)\dot{u}_{\omega}^{(\textsc{in})*}(t)-u_{\omega}^{(\textsc{in})*}(t)\dot{u}_{\omega}^{(\textsc{in})\vphantom{*}}(t)&=i\,,&\bigl(N_{\bm{k}}^{(\textsc{out})}+\frac{1}{2}\bigr)^{2}-\lvert C_{\bm{k}}^{(\textsc{out})}\rvert^{2}&=\frac{1}{4}
\end{align}
for the current setting, we find
\begin{equation}
	\mathcal{W}_{\bm{k}}^{(i)}(\Sigma_{\bm{k}}^{(i)},P_{\bm{k}}^{(i)};t)=\frac{1}{\pi}\,\exp\Bigl\{-2\Bigl[2\bigl(N_{\bm{k}}^{\textsc{(out)}}+\frac{1}{2}\bigr)D_{\bm{k}}^{(i)}D_{\bm{k}}^{(i)*}+C_{\bm{k}}^{\textsc{(out)}\vphantom{*}}D_{\bm{k}}^{(i)2}+C_{\bm{k}}^{\textsc{(out)}*}D_{\bm{k}}^{(i)*2}\Bigr]\Bigr\}\,,\label{E:dgnsjd}
\end{equation}
where $D_{\bm{k}}^{(i)}=P_{\bm{k}}^{(i)}u_{\omega}^{(\textsc{in})}-\Sigma_{\bm{k}}^{(i)}\dot{u}_{\omega}^{(\textsc{in})}$. We have expressed the coefficients in the exponent of the Gaussian-state Wigner function in terms of $N_{\bm{k}}$ and $C_{\bm{k}}$.

Eq.~\eqref{E:dgnsjd}   is not quite what we want because it is written in terms of the real and the imaginary parts of the field modes. Rather we would like to write it as a function of the field modes $\phi_{\bm{k}}$ and the conjugate momenta $\pi_{\bm{k}}$. Inspired by \eqref{E:rkjgb1} and \eqref{E:rkjgb2}
\begin{align}
	\hat{\phi}_{\bm{k}}^{(1)}&=\frac{1}{\sqrt{2}}\bigl(\hat{\phi}_{\bm{k}}^{\vphantom{\dagger}}+\hat{\phi}_{\bm{k}}^{\dagger}\bigr)\,,&\hat{\phi}_{\bm{k}}^{(2)}&=\frac{1}{i\sqrt{2}}\bigl(\hat{\phi}_{\bm{k}}^{\vphantom{\dagger}}-\hat{\phi}_{\bm{k}}^{\dagger}\bigr)\,,\label{E:fgjdh1}\\
	\hat{p}_{\bm{k}}^{(1)}&=\frac{1}{\sqrt{2}}\bigl(\hat{\pi}_{\bm{k}}^{\dagger}+\hat{\pi}_{\bm{k}}^{\vphantom{\dagger}}\bigr)\,,&\hat{p}_{\bm{k}}^{(2)}&=\frac{1}{i\sqrt{2}}\bigl(\hat{\pi}_{\bm{k}}^{\dagger}-\hat{\phi}_{\bm{k}}^{\vphantom{\dagger}}\bigr)\,,\label{E:fgjdh2}
\end{align}
we will combine the Wigner functions $\mathcal{W}_{\bm{k}}^{(1)}(\Sigma_{\bm{k}}^{(1)},P_{\bm{k}}^{(1)};t)$ and $\mathcal{W}_{\bm{k}}^{(2)}(\Sigma_{\bm{k}}^{(2)},P_{\bm{k}}^{(2)};t)$ to form the Wigner function for the modes $\pm\bm{k}$
\begin{align}\label{E:rusbdhssd}
	\mathcal{W}_{\pm\bm{k}}(\Sigma_{+\bm{k}},\Pi_{+\bm{k}};\Sigma_{-\bm{k}},\Pi_{-\bm{k}};t)&=\mathcal{W}_{\bm{k}}^{(1)}(\Sigma_{\bm{k}}^{(1)},P_{\bm{k}}^{(1)};t)\mathcal{W}_{\bm{k}}^{(2)}(\Sigma_{\bm{k}}^{(2)},P_{\bm{k}}^{(2)};t)\notag\\
	&=\biggl(\frac{1}{\pi}\biggr)^{2}\exp\Bigl\{-2\Bigl[2\bigl(N_{\bm{k}}^{\textsc{(out)}}+\frac{1}{2}\bigr)\bigl(D_{\bm{k}}^{(1)}D_{\bm{k}}^{(1)*}+D_{\bm{k}}^{(2)}D_{\bm{k}}^{(2)*}\bigr)\Bigr.\biggr.\\
	&\qquad\qquad\qquad\quad+\biggl.\Bigl.C_{\bm{k}}^{\textsc{(out)}\vphantom{*}}\bigl(D_{\bm{k}}^{(1)2}+D_{\bm{k}}^{(2)2}\bigr)+C_{\bm{k}}^{\textsc{(out)}*}\bigl(D_{\bm{k}}^{(1)*2}+D_{\bm{k}}^{(2)*2}\bigr)\Bigr]\Bigr\}\,.\notag
\end{align}
Observe that we can write $D_{\bm{k}}^{(1)}D_{\bm{k}}^{(1)*}+D_{\bm{k}}^{(2)}D_{\bm{k}}^{(2)*}$ into
\begin{align*}
	D_{\bm{k}}^{(1)}D_{\bm{k}}^{(1)*}+D_{\bm{k}}^{(2)}D_{\bm{k}}^{(2)*}=\frac{1}{2}\Bigl(D_{\bm{k}}^{(1)}+i\,D_{\bm{k}}^{(2)}\Bigr)\Bigl(D_{\bm{k}}^{(1)*}-i\,D_{\bm{k}}^{(2)*}\Bigr)+\frac{1}{2}\Bigl(D_{\bm{k}}^{(1)}-i\,D_{\bm{k}}^{(2)}\Bigr)\Bigl(D_{\bm{k}}^{(1)*}+i\,D_{\bm{k}}^{(2)*}\Bigr)\,,
\end{align*}
and then introduce a new set of variables $D_{\pm\bm{k}}$,
\begin{align}
	\frac{1}{\sqrt{2}}\Bigl(D_{\bm{k}}^{(1)}+i\,D_{\bm{k}}^{(2)}\Bigr)&=\Pi_{-\bm{k}}\,u_{\omega}^{(\textsc{in})\vphantom{*}}-\Sigma_{+\bm{k}}\,\dot{u}_{\omega}^{(\textsc{in})\vphantom{*}}\equiv D_{+\bm{k}}\,,\\
	\frac{1}{\sqrt{2}}\Bigl(D_{\bm{k}}^{(1)}-i\,D_{\bm{k}}^{(2)}\Bigr)&=\Pi_{+\bm{k}}\,u_{\omega}^{(\textsc{in})\vphantom{*}}-\Sigma_{-\bm{k}}\,\dot{u}_{\omega}^{(\textsc{in})\vphantom{*}}\equiv D_{-\bm{k}}\,,
\end{align}
with
\begin{align}
	\Pi_{\bm{k}}&=\frac{1}{\sqrt{2}}\bigl(P_{\bm{k}}^{(1)}-i\,P_{\bm{k}}^{(2)}\bigr)\,,&\Sigma_{\bm{k}}&=\frac{1}{\sqrt{2}}\bigl(\Sigma_{\bm{k}}^{(1)}+i\,\Sigma_{\bm{k}}^{(2)}\bigr)\,.
\end{align}
Thus we arrive at
\begin{align}
	D_{\bm{k}}^{(1)}D_{\bm{k}}^{(1)*}+D_{\bm{k}}^{(2)}D_{\bm{k}}^{(2)*}=D_{+\bm{k}}^{\vphantom{*}}D_{+\bm{k}}^{*}+D_{-\bm{k}}^{\vphantom{*}}D_{-\bm{k}}^{*}\,.
\end{align}
Note that in general $D_{+\bm{k}}^{*}\neq D_{-\bm{k}}^{\vphantom{*}}$. Similarly we find
\begin{align}
	D_{\bm{k}}^{(1)2}+D_{\bm{k}}^{(2)2}&=\Bigl(D_{\bm{k}}^{(1)}+i\,D_{\bm{k}}^{(2)}\Bigr)\Bigl(D_{\bm{k}}^{(1)}-i\,D_{\bm{k}}^{(2)}\Bigr)=2D_{+\bm{k}}^{\vphantom{*}}D_{-\bm{k}}^{\vphantom{*}}\,,\\
	D_{\bm{k}}^{(1)*2}+D_{\bm{k}}^{(2)*2}&=2D_{+\bm{k}}^{*}D_{-\bm{k}}^{*}\,.
\end{align}
Therefore the Wigner function \eqref{E:rusbdhssd} becomes
\begin{align}\label{E:tjdkfgsd}
	&\quad\mathcal{W}_{\pm\bm{k}}(\Sigma_{+\bm{k}},\Pi_{+\bm{k}};\Sigma_{-\bm{k}},\Pi_{-\bm{k}};t)\\
	&=\biggl(\frac{1}{\pi}\biggr)^{2}\exp\Bigl\{-4\Bigl[\bigl(N_{\bm{k}}^{\textsc{(out)}}+\frac{1}{2}\bigr)\bigl(D_{+\bm{k}}^{\vphantom{*}}D_{+\bm{k}}^{*}+D_{-\bm{k}}^{\vphantom{*}}D_{-\bm{k}}^{*}\bigr)+C_{\bm{k}}^{\textsc{(out)}\vphantom{*}}D_{+\bm{k}}^{\vphantom{*}}D_{-\bm{k}}^{\vphantom{*}}+C_{\bm{k}}^{\textsc{(out)}*}D_{+\bm{k}}^{*}D_{-\bm{k}}^{*}\Bigr]\Bigr\}\notag\,.
\end{align}
We arrive at a Winger function in terms of a new set of variables $D_{\pm\bm{k}}$, a superposition of the field mode $\Sigma_{\bm{k}}$ and its conjugate momentum $\Pi_{\bm{k}}$, but they have an unambiguous sense of parity due to $D_{+\bm{k}}^{*}\neq D_{-\bm{k}}^{\vphantom{*}}$, unlike $\Sigma_{\bm{k}}$ and $\Pi_{\bm{k}}$. In contrast, the latter set of variables in mode $+\bm{k}$ can be related to their counterparts in the $-\bm{k}$ mode by complex conjugation.

This nice feature allows us to unambiguously identify the effect of coarse-graining the field. Suppose we lose track of the information regarding the $-\bm{k}$ modes of the created particles, so we coarse-grain out the contribution of the $-\bm{k}$ mode in the Wigner equation \eqref{E:tjdkfgsd}. Then we obtain
\begin{align}
	\mathcal{W}_{\bm{k}}(\Sigma_{\bm{k}},\Pi_{\bm{k}};t)&=\int\!dD_{-\bm{k}}^{\vphantom{*}}dD_{-\bm{k}}^{*}\;\mathcal{W}_{\pm\bm{k}}(\Sigma_{+\bm{k}},\Pi_{+\bm{k}};\Sigma_{-\bm{k}},\Pi_{-\bm{k}};t)\notag\\
	&=\frac{1}{2\pi(N_{\bm{k}}^{\textsc{(out)}}+1/2)}\exp\biggl\{-\frac{4}{(N_{\bm{k}}^{\textsc{(out)}}+1/2)}\,D_{\bm{k}}^{\vphantom{*}}D_{\bm{k}}^{*}\Bigl[\Bigl(N_{\bm{k}}^{\textsc{(out)}}+\frac{1}{2}\Bigr)-\lvert C_{\bm{k}}^{\textsc{(out)}}\rvert^{2}\Bigr]\biggr\}\notag\\
	&=\frac{1}{2\pi(N_{\bm{k}}^{\textsc{(out)}}+1/2)}\exp\biggl\{-\frac{1}{(N_{\bm{k}}^{\textsc{(out)}}+1/2)}\,D_{\bm{k}}^{\vphantom{*}}D_{\bm{k}}^{*}\biggr\}\,.\label{E:rljnbdkhgf}
\end{align}
The resulting reduced Wigner function is independent of $C_{\bm{k}}^{(\textsc{out})}$. It implies that the correlation between the particles of $\pm\bm{k}$ is completely lost. Thus we expect the corresponding entropy will be a function of $N_{\bm{k}}^{(\textsc{out})}$ only.

To better understand what \eqref{E:rljnbdkhgf} tells, and more specifically, the meaning of the variables $D_{\bm{k}}$, we first note that \eqref{E:rljnbdkhgf} is essentially \eqref{E:dgnsjd} with $C_{\bm{k}}^{(\textsc{out})}=0$ and a rescaling of the variable $D_{\bm{k}}^{(i)}$ by
\begin{equation}
	D_{\bm{k}}^{(i)}\mapsto\frac{1}{2N_{\bm{k}}^{\textsc{(out)}}+1}\,D_{\bm{k}}^{(i)}
\end{equation}
in \eqref{E:dgnsjd}, so we expect the corresponding entropy will be given by \eqref{E:fgjhdfgdx} with $C_{\bm{k}}^{(\textsc{out})}=0$ in $\lambda_{\bm{k}}$, that is, 
\begin{align}\label{E:fgvdjf}
	\mathcal{S}_{\bm{k}}&=\bigl(N_{\bm{k}}+1\bigr)\,\ln\bigl(N_{\bm{k}}+1\bigr)-N_{\bm{k}}\,\ln N_{\bm{k}}\,,
\end{align}
with a corresponding particle number $N_{\bm{k}}$ for each $\bm{k}>0$ mode.  {Thus the previous arguments lead to the conclusion that the field will have entropy production associated with the creation of particle pairs only when we lose track of some information embodied in the field -- in this case, all the information about one partner of the pair. The nonzero entropy arises from the lack of complete information of the field, not from the creation of field quanta. Moreover, since the state of each field mode before coarse-graining is pure, the nonzero value of the entropy \eqref{E:fgvdjf} also indicates that the particle pair is entangled~\cite{LCH10}. Now it is clear that what is deeper inside correlation or coherence between the particle pair is indeed quantum entanglement between them.}

To find out the meaning of $D_{\bm{k}}$, we first write the field by the out-mode functions, 
\begin{align}
	\Sigma_{\bm{k}}&=\mathfrak{a}_{+\bm{k}}^{\vphantom{*}}\,u_{\omega}^{(\textsc{out})}+\mathfrak{a}_{-\bm{k}}^{*}\,u_{\omega}^{(\textsc{out})*}\,,&\Pi_{\bm{k}}&=\mathfrak{a}_{-\bm{k}}^{\vphantom{*}}\,\dot{u}_{\omega}^{(\textsc{out})}+\mathfrak{a}_{+\bm{k}}^{*}\,\dot{u}_{\omega}^{(\textsc{out})*}\,,
\end{align}
where $\mathfrak{a}_{\bm{k}}$ are the expansion coefficient, the $c$-number counterpart of the in-mode operator $\hat{a}_{\bm{k}}$. Then since the in-mode functions and the out-mode functions are related by
\begin{equation}
	u_{\omega}^{(\textsc{out})}(t)=\alpha_{\bm{k}}\,u_{\omega}^{(\textsc{in})}+\beta_{\bm{k}}\,u_{\omega}^{(\textsc{in})*}\,,
\end{equation}
we find
\begin{align}
	D_{\bm{k}}&=\Pi_{\bm{k}}^{*}\,u_{\omega}^{(\textsc{in})\vphantom{*}}-\Sigma_{\bm{k}}^{\vphantom{*}}\,\dot{u}_{\omega}^{(\textsc{in})\vphantom{*}}=\bigl(\alpha_{\bm{k}}^{*}\mathfrak{a}_{-\bm{k}}^{*}+\beta_{\bm{k}}^{\vphantom{*}}\mathfrak{a}_{+\bm{k}}^{\vphantom{*}}\bigr)\bigl(u_{\omega}^{(\textsc{in})}\dot{u}_{\omega}^{(\textsc{in})*}-u_{\omega}^{(\textsc{in})*}\dot{u}_{\omega}^{(\textsc{in})}\bigr)\,.
\end{align}
The second pair of parentheses give an $i$ due to the Wronskian condition of the mode functions, while the expressions inside the first pair of parentheses is worth further investigation. From \eqref{E:ghjrbdfdf}, we have
\begin{equation}
	\mathfrak{b}_{-\bm{k}}^{*}=\alpha_{-\bm{k}}^{*}\,\mathfrak{a}_{-\bm{k}}^{*}+\beta_{+\bm{k}}^{\vphantom{*}}\,\mathfrak{a}_{\bm{k}}^{\vphantom{*}}=\alpha_{\omega}^{*}\,\mathfrak{a}_{-\bm{k}}^{*}+\beta_{\omega}^{\vphantom{*}}\,\mathfrak{a}_{\bm{k}}^{\vphantom{*}}\,,\label{E:ffkdgs}
\end{equation}
because for our configuration the Bogoliubov coefficients are only functions of $\lvert\bm{k}\rvert$. Thus we  {find that the variable $D_{\bm{k}}^{\vphantom{*}}$ actually is}
\begin{equation}
	D_{\bm{k}}^{\vphantom{*}}=i\,\mathfrak{b}_{-\bm{k}}^{*}\,,
\end{equation}
{that is, expansion coefficient of $\Sigma_{\bm{k}}$ with respect to the IN mode}, so the Wigner function \eqref{E:rljnbdkhgf} now becomes
\begin{equation}
	\mathcal{W}_{\bm{k}}(t)=\frac{1}{2\pi(N_{\bm{k}}^{\textsc{(out)}}+1/2)}\exp\biggl\{-\frac{1}{(N_{\bm{k}}^{\textsc{(out)}}+1/2)}\,\mathfrak{b}_{-\bm{k}}^{\vphantom{*}}\mathfrak{b}_{-\bm{k}}^{*}\biggr\}\,.
\end{equation}
The labeling of the Wigner function can be a little awkward because it is expanded by the variables of the $-\bm{k}$ mode.

It is interesting to note that once we sever the correlation between the $+\bm{k}$ and the $-\bm{k}$ modes, the resulting Wigner function seems to be ``instantaneously'' stationary in the out-variables. However, it actually is not,  because the associated coefficients are all time-dependent.

In summary,    for the parametric processes of  interest here, the field evolves unitarily, so even though particles are produced, there is no entropy production. {Quantum entanglement} exists between the created particle pairs. However, if  some information of the field is lost, then the entropy of the field can increase even if the field is initially in a pure state.  In particular,   an entropy of the form \eqref{E:fgvdjf} is obtained if information is lost coming from  one of the  particle pairs produced for each mode.

\subsection{Wigner function of the quantum field}

It has been been shown that the statistical properties of the classical stochastic field can be described by a classical distribution functional of the field variable and its conjugate momentum over an infinite-dimensional phase space spanned by the canonical pair~\cite{BMP}. However it is not explained in that context how a classical stochastic field can act as a full delegate of a quantum field.  It will be more desirable to have a full quantum description for the statistical process of the field. Here we provide such a formulation based on \eqref{E:kfghbs}. 

We will construct the Wigner function in terms of the full field $\phi(\bm{x},t)$, instead of its modes $\phi_{\bm{k}}(t)$. From \eqref{E:kfghbs}, we have the Wigner function of all modes give by
\begin{align}\label{E:ieuere}
	&\quad\prod_{\bm{k},i}\mathcal{W}_{\bm{k}}^{(i)}(\phi_{\bm{k}}^{(i)},p_{\bm{k}}^{(i)};t)\\
	&=\biggl(\prod_{\bm{k}>0}\mathcal{N}_{\bm{k}}^{(w)2}\biggr)\exp\biggl\{-\frac{1}{2}\sum_{\bm{k}>0}\biggl[\mathsf{A}_{\bm{k}}^{\vphantom{2}}\Bigl(\phi_{\bm{k}}^{(1)2}+\phi_{\bm{k}}^{(2)2}\Bigr)+\mathsf{B}_{\bm{k}}^{\vphantom{2}}\Bigl(p_{\bm{k}}^{(1)2}+p_{\bm{k}}^{(2)2}\Bigr)+2\mathsf{C}_{\bm{k}}^{\vphantom{2}}\Bigl(\phi_{\bm{k}}^{(1)}p_{\bm{k}}^{(1)}+\phi_{\bm{k}}^{(2)}p_{\bm{k}}^{(2)}\Bigr)\biggr]\biggr\}\,,\notag
\end{align}
where we note that $\phi_{\bm{k}}^{(i)}$ and $p_{\bm{k}}^{(i)}$ are $c$-number, and
\begin{align}
	\mathcal{N}^{(w)}_{\bm{k}}&=\frac{1}{2\pi\sqrt{\mathsf{a}_{\bm{k}}^{\vphantom{2}}\mathsf{b}_{\bm{k}}^{\vphantom{2}}-\mathsf{c}_{\bm{k}}^{2}}}\,,&\mathsf{A}_{\bm{k}}^{\vphantom{2}}&=\frac{\mathsf{a}_{\bm{k}}}{\mathsf{a}_{\bm{k}}^{\vphantom{2}}\mathsf{b}_{\bm{k}}^{\vphantom{2}}-\mathsf{c}_{\bm{k}}^{2}}\,,&\mathsf{B}_{\bm{k}}^{\vphantom{2}}&=\frac{\mathsf{b}_{\bm{k}}}{\mathsf{a}_{\bm{k}}^{\vphantom{2}}\mathsf{b}_{\bm{k}}^{\vphantom{2}}-\mathsf{c}_{\bm{k}}^{2}}\,,&\mathsf{C}_{\bm{k}}^{\vphantom{2}}&=-\frac{\mathsf{c}_{\bm{k}}}{\mathsf{a}_{\bm{k}}^{\vphantom{2}}\mathsf{b}_{\bm{k}}^{\vphantom{2}}-\mathsf{c}_{\bm{k}}^{2}}\,.\label{E:dlfnbksd}
\end{align}
Since following \eqref{E:rkjgb1} and \eqref{E:rkjgb2}, we immediately have the $c$-number counterparts of \eqref{E:fgjdh1} and \eqref{E:fgjdh2} given by
\begin{align}
	\phi_{\bm{k}}^{(1)}&=\frac{1}{\sqrt{2}}\bigl(\phi_{\bm{k}}^{\vphantom{*}}+\phi_{\bm{k}}^{*}\bigr)\,,&\phi_{\bm{k}}^{(2)}&=\frac{1}{i\sqrt{2}}\bigl(\phi_{\bm{k}}^{\vphantom{*}}-\phi_{\bm{k}}^{*}\bigr)\,,\\
	p_{\bm{k}}^{(1)}&=\frac{1}{\sqrt{2}}\bigl(\pi_{\bm{k}}^{*}+\pi_{\bm{k}}^{\vphantom{*}}\bigr)\,,&p_{\bm{k}}^{(2)}&=\frac{1}{i\sqrt{2}}\bigl(\pi_{\bm{k}}^{*}-\phi_{\bm{k}}^{\vphantom{*}}\bigr)\,,
\end{align}
so that we obtain
\begin{align}
	\phi_{\bm{k}}^{(1)2}+\phi_{\bm{k}}^{(2)2}&=2\phi_{\bm{k}}^{\vphantom{*}}\phi_{\bm{k}}^{*}=2\phi_{-\bm{k}}^{\vphantom{*}}\phi_{-\bm{k}}^{*}=2\phi_{\bm{k}}^{\vphantom{*}}\phi_{-\bm{k}}^{\vphantom{*}}=\phi_{\bm{k}}^{\vphantom{*}}\phi_{\bm{k}}^{*}+\phi_{-\bm{k}}^{\vphantom{*}}\phi_{-\bm{k}}^{*}\,,\\
	p_{\bm{k}}^{(1)2}+p_{\bm{k}}^{(2)2}&=2\pi_{\bm{k}}^{\vphantom{*}}\pi_{\bm{k}}^{*}\,,\\
	\phi_{\bm{k}}^{(1)}p_{\bm{k}}^{(1)}+\phi_{\bm{k}}^{(2)}p_{\bm{k}}^{(2)}&=\phi_{\bm{k}}^{\vphantom{*}}\pi_{\bm{k}}^{\vphantom{*}}+\phi_{\bm{k}}^{*}\pi_{\bm{k}}^{*}=\phi_{\bm{k}}^{\vphantom{*}}\pi_{\bm{k}}^{\vphantom{*}}+\phi_{-\bm{k}}^{\vphantom{*}}\pi_{-\bm{k}}^{\vphantom{*}}\,.
\end{align}
If the initial state is stationary, then from \eqref{E:eirh1} and \eqref{E:eirh2} we find
\begin{align}
	\mathsf{A}_{+\bm{k}}^{\vphantom{2}}&=\mathsf{A}_{-\bm{k}}^{\vphantom{2}}\,,&\mathsf{B}_{+\bm{k}}^{\vphantom{2}}&=\mathsf{B}_{-\bm{k}}^{\vphantom{2}}\,,&\mathsf{C}_{+\bm{k}}^{\vphantom{2}}&=\mathsf{C}_{-\bm{k}}^{\vphantom{2}}\,.
\end{align}
and then we obtain
\begin{align}
	\prod_{\bm{k},i}\mathcal{W}_{\bm{k}}^{(i)}(\phi_{\bm{k}}^{(i)},p_{\bm{k}}^{(i)};t)&=\biggl(\prod_{\bm{k}}\mathcal{N}^{(w)}_{\bm{k}}\biggr)\exp\biggl\{-\frac{1}{2}\sum_{\bm{k}}\biggl[\mathsf{A}_{\bm{k}}^{\vphantom{2}}\phi_{\bm{k}}^{\vphantom{*}}\phi_{\bm{k}}^{*}+\mathsf{B}_{\bm{k}}^{\vphantom{2}}\pi_{\bm{k}}^{\vphantom{*}}\pi_{\bm{k}}^{*}+2\mathsf{C}_{\bm{k}}^{\vphantom{2}}\phi_{\bm{k}}^{\vphantom{*}}\pi_{\bm{k}}^{\vphantom{*}}\biggr]\biggr\}.
\end{align}
Observe that the functional ``product'' of the stationary functions
\begin{equation}
	f(\bm{x}-\bm{z})=\int_{-\infty}^{\infty}\!d^{3}y\;g(\bm{x}-\bm{y})h(\bm{y}-\bm{z})
\end{equation}
can be written as
\begin{align}
	f(\bm{x}-\bm{z})=\int_{-\infty}^{\infty}\!d^{3}y\;g(\bm{x}-\bm{y})h(\bm{y}-\bm{z})&=\int\!\!\frac{d^{3}k}{(2\pi)^{3}}\;g_{\bm{k}}h_{\bm{k}}\,e^{+i\bm{k}\cdot(\bm{x}-\bm{z})}\,,
\end{align}
and thus $f_{\bm{k}}=g_{\bm{k}}h_{\bm{k}}$, where the spatial Fourier transformation is defined by
\begin{equation}
	f(\bm{x})=\int\!\!\frac{d^{3}k}{(2\pi)^{3}}\,f_{\bm{k}}\,e^{+i\bm{k}\cdot\bm{x}}\,.
\end{equation}
It implies that the inverse in this sense is given by
\begin{align}
	\delta^{(3)}(\bm{x}-\bm{z})&=\int_{-\infty}^{\infty}\!d^{3}y\;f(\bm{x}-\bm{y})g(\bm{y}-\bm{z})\,,&&\Rightarrow&1&=f_{\bm{k}}g_{\bm{k}}\,,&&\Rightarrow&f_{\bm{k}}^{-1}&=\frac{1}{g_{\bm{k}}}\,.
\end{align}
That is, in this context, the notion of the inverse $g(\bm{x}-\bm{y})=f^{-1}(\bm{x}-\bm{y})$ of a function $f(\bm{x}-\bm{y})$ does not mean the usual inverse function, nor $1/f(\bm{x}-\bm{y})$. We then find that their Fourier transforms obey the standard rule of the multiplication.

Therefore, the convolution relation for $\phi(\bm{x},t)$ reduces to
\begin{align}
	\int_{-\infty}^{\infty}\!d^{3}x\!\int_{-\infty}^{\infty}\!d^{3}y\;\phi(\bm{x},t)\mathsf{A}(\bm{x}-\bm{y};t)\phi(\bm{y},t)&=\int\!\!\frac{d^{3}k}{(2\pi)^{3}}\;\phi_{\bm{k}}^{\vphantom{*}}(t)\mathsf{A}_{\bm{k}}(t)\phi_{\bm{k}}^{*}(t)\,,
\end{align}
On the other hand, the convolution relations are different for $\pi(\bm{x},t)$ due to \eqref{E:qqzkgjbe}. For example we find
\begin{align}
	\int_{-\infty}^{\infty}\!d^{3}x\!\int_{-\infty}^{\infty}\!d^{3}y\;\phi(\bm{x},t)\mathsf{C}(\bm{x}-\bm{y};t)\pi(\bm{y},t)&=\int\!\!\frac{d^{3}k}{(2\pi)^{3}}\;\phi_{\bm{k}}^{\vphantom{*}}(t)\mathsf{C}_{-\bm{k}}^{\vphantom{*}}(t)\pi_{\bm{k}}^{\vphantom{*}}(t)\int\!\!\frac{d^{3}k}{(2\pi)^{3}}\;\phi_{\bm{k}}^{\vphantom{*}}(t)\mathsf{C}_{\bm{k}}^{\vphantom{*}}(t)\pi_{\bm{k}}^{\vphantom{*}}(t)\,,
\end{align}
and
\begin{align}
	\int_{-\infty}^{\infty}\!d^{3}x\!\int_{-\infty}^{\infty}\!d^{3}y\;\pi(\bm{x},t)\mathsf{B}(\bm{x}-\bm{y};t)\pi(\bm{y},t)&=\int\!\!\frac{d^{3}k}{(2\pi)^{3}}\;\pi^{\vphantom{*}}_{\bm{k}}(t)\mathsf{B}_{\bm{k}}^{\vphantom{*}}(t)\pi_{\bm{k}}^{*}(t)\,.
\end{align}
This implies that the Wigner function of the full field will be given by
\begin{align}
	\mathcal{W}(\phi,\pi;t)&=\prod_{\bm{k},i}\mathcal{W}_{\bm{k}}^{(i)}(\phi_{\bm{k}}^{(i)},p_{\bm{k}}^{(i)};t)\notag\\
	&=\mathcal{N}^{(w)}\,\exp\biggl\{-\frac{1}{2}\int_{-\infty}^{\infty}\!d^{3}x\!\int_{-\infty}^{\infty}\!d^{3}y\biggl[\phi(\bm{x},t)\mathsf{A}(\bm{x}-\bm{y};t)\phi(\bm{y},t)+\pi(\bm{x},t)\mathsf{B}(\bm{x}-\bm{y};t)\pi(\bm{y},t)\biggr.\biggr.\notag\\
	&\qquad\qquad\qquad\qquad\qquad\qquad\qquad\qquad+\biggl.\biggl.2\phi(\bm{x},t)\mathsf{C}(\bm{x}-\bm{y};t)\pi(\bm{y},t)\biggr]\biggr\}\,,\label{E:risbdfd}
\end{align}
where we have used the fact that the covariance matrix elements
\begin{align*}
	\mathsf{b}_{\bm{k}}&=\frac{1}{2}\langle\bigl\{\hat{\phi}_{\bm{k}}^{\vphantom{\dagger}},\,\hat{\phi}_{\bm{k}}^{\dagger}\bigr\}\rangle\,,&\mathsf{a}_{\bm{k}}&=\frac{1}{2}\langle\bigl\{\hat{\pi}_{\bm{k}}^{\vphantom{\dagger}},\,\hat{\pi}_{\bm{k}}^{\dagger}\bigr\}\rangle\,,&\mathsf{c}_{\bm{k}}&=\frac{1}{2}\Bigl[\frac{1}{2}\langle\bigl\{\hat{\phi}_{\bm{k}},\,\hat{\pi}_{\bm{k}}\bigr\}\rangle+\frac{1}{2}\langle\bigl\{\hat{\phi}_{\bm{k}}^{\dagger},\,\hat{\pi}_{\bm{k}}^{\dagger}\bigr\}\rangle\Bigr]\,,
\end{align*}
are all real. The normalization constant $\mathcal{N}^{(w)}$ will be determined by the functional integral
\begin{equation}
	\int\!\mathcal{D}\phi\mathcal{D}\pi\;\mathcal{W}(\phi,\pi;t)=1\,.
\end{equation}
The ``inverse'' of the coefficient functions $\mathsf{A}(\bm{x}-\bm{y};t)$, $\mathsf{B}(\bm{x}-\bm{y};t)$ and $\mathsf{C}(\bm{x}-\bm{y};t)$ take the simpler forms than the coefficient functions themselves because
\begin{align}
	\mathsf{A}_{\bm{k}}^{-1}&=\mathsf{b}_{\bm{k}}-\mathsf{c}_{\bm{k}}^{\vphantom{-1}}\mathsf{a}_{\bm{k}}^{-1}\mathsf{c}_{\bm{k}}^{\vphantom{-1}}\,,&\mathsf{B}_{\bm{k}}^{-1}&=\mathsf{a}_{\bm{k}}-\mathsf{c}_{\bm{k}}^{\vphantom{-1}}\mathsf{b}_{\bm{k}}^{-1}\mathsf{c}_{\bm{k}}^{\vphantom{-1}}\,,&\mathsf{C}_{\bm{k}}^{-1}&=\mathsf{b}_{\bm{k}}-\mathsf{a}_{\bm{k}}^{\vphantom{-1}}\mathsf{c}_{\bm{k}}^{-1}\mathsf{b}_{\bm{k}}^{\vphantom{-1}}\,.
\end{align}
Then we find
\begin{align}
	\mathsf{A}^{-1}(\bm{x}-\bm{y};t)&=\int\!\!\frac{d^{3}k}{(2\pi)^{3}}\;\mathsf{A}_{\bm{k}}^{-1}e^{+i\bm{k}\cdot(\bm{x}-\bm{y})}=\int\!\!\frac{d^{3}k}{(2\pi)^{3}}\;\Bigl[\mathsf{b}_{\bm{k}}-\mathsf{c}_{\bm{k}}^{\vphantom{1}}\mathsf{a}_{\bm{k}}^{-1}\mathsf{c}_{\bm{k}}^{\vphantom{1}}\Bigr]e^{+i\bm{k}\cdot(\bm{x}-\bm{y})}\notag\\
	&=\mathsf{b}(\bm{x}-\bm{y};t)-\int\!d^{3}ud^{3}v\;\mathsf{c}(\bm{x}-\bm{u};t)\,\mathsf{a}^{-1}(\bm{u}-\bm{v};t)\,\mathsf{c}(\bm{v}-\bm{y};t)\,,\\
	\mathsf{B}^{-1}(\bm{x}-\bm{y};t)&=\mathsf{a}(\bm{x}-\bm{y};t)-\int\!d^{3}ud^{3}v\;\mathsf{c}(\bm{x}-\bm{u};t)\,\mathsf{b}^{-1}(\bm{u}-\bm{v};t)\,\mathsf{c}(\bm{v}-\bm{y};t)\,,\\
	\mathsf{C}^{-1}(\bm{x}-\bm{y};t)&=\mathsf{c}(\bm{x}-\bm{y};t)-\int\!d^{3}ud^{3}v\;\mathsf{a}(\bm{x}-\bm{u};t)\,\mathsf{c}^{-1}(\bm{u}-\bm{v};t)\,\mathsf{b}(\bm{v}-\bm{y};t)\,,
\end{align}
in the sense of functional multiplication. In deriving the above results we have used
\begin{align*}
	\mathsf{b}(\bm{x}-\bm{y};t)=\int\!\!\frac{d^{3}k}{(2\pi)^{3}}\;\mathsf{b}_{\bm{k}}(t)\,e^{+i\bm{k}\cdot(\bm{x}-\bm{y})}&=\int\!\!\frac{d^{3}k}{(2\pi)^{3}}\;\frac{1}{2}\langle\bigl\{\hat{\phi}_{\bm{k}}^{\vphantom{\dagger}}(t),\,\hat{\phi}_{\bm{k}}^{\dagger}(t)\bigr\}\rangle\,e^{+i\bm{k}\cdot(\bm{x}-\bm{y})}=\frac{1}{2}\langle\bigl\{\hat{\phi}(\bm{x},t),\,\hat{\phi}(\bm{y},t)\bigr\}\rangle\,,
\end{align*}
due to \eqref{E:ngsjggh1}, and similarly
\begin{equation*}
	\mathsf{a}(\bm{x}-\bm{y};t)=\int\!\!\frac{d^{3}k}{(2\pi)^{3}}\;\mathsf{a}_{\bm{k}}(t)\,e^{+i\bm{k}\cdot(\bm{x}-\bm{y})}=\frac{1}{2}\langle\bigl\{\hat{\pi}(\bm{x},t),\,\hat{\pi}(\bm{y},t)\bigr\}\rangle\,,
\end{equation*}
and
\begin{align}
	\mathsf{c}(\bm{x}-\bm{y};t)=\int\!\!\frac{d^{3}k}{(2\pi)^{3}}\;\mathsf{c}_{\bm{k}}(t)\,e^{+i\bm{k}\cdot(\bm{x}-\bm{y})}&=\int\!\!\frac{d^{3}k}{(2\pi)^{3}}\;\frac{1}{2}\Bigl[\frac{1}{2}\langle\bigl\{\hat{\phi}_{\bm{k}},\,\hat{\pi}_{\bm{k}}\bigr\}\rangle+\frac{1}{2}\langle\bigl\{\hat{\phi}_{\bm{k}}^{\dagger},\,\hat{\pi}_{\bm{k}}^{\dagger}\bigr\}\rangle\Bigr]\,e^{+i\bm{k}\cdot(\bm{x}-\bm{y})}\notag\\
	&=\frac{1}{2}\,\langle\bigl\{\hat{\phi}(\bm{x},t),\,\hat{\pi}(\bm{y},t)\bigr\}\rangle\,,
\end{align}
owing to \eqref{E:ngsjggh3}. Again we remind that $\mathsf{A}^{-1}(\bm{x}-\bm{y};t)$ does not represent the reciprocal of $\mathsf{A}(\bm{x}-\bm{y};t)$ per se.

We therefore see that the distribution function of the classical stochastic field happens to take exactly the same form as the Wigner function \eqref{E:risbdfd} of the quantum scalar field, but our derivation is based on quantum field theory. Even though both formally look identical, the frameworks they are base upon are drastically distinct. Other than the lurid difference of quantum non-commutativity, there are a few additional subtleties. The Wigner function is known not to be always positive-definite, in contrast to the classical probability distribution. Classical field theory cannot account for quantum entanglement,  in which the nature of quantum state plays an important role. Moreover, the Shannon entropy associated with  classical distributions has the property of monotonicity. That is, given the combined systems $A$ and $B$, we have $\mathcal{S}_{s}[A+B]\geq\max\{\mathcal{S}_{s}[A],\mathcal{S}_{s}[B]\}$ for the Shannon entropy, where $\mathcal{S}_{s}[A]$ denotes the Shannon entropy of the subsystem $A$. On the other hand, for the von Neumann entropy $\mathcal{S}$, the closest property to the classical monotonicity is the theorem of Araki-Lieb  triangle inequality
\begin{equation}
	\lvert \mathcal{S}[A]-\mathcal{S}[B]\rvert\leq \mathcal{S}[A+B]\leq \mathcal{S}[A]+\mathcal{S}[B]\,.
\end{equation}
The second inequalities is known as the subadditivity inequality for von Neumann entropy, and holds with equality if and only if systems $A$ and $B$ are uncorrelated, that is, $\varrho^{AB}=\varrho^{A}\otimes\varrho^{B}$. In the case we discussed earlier, let the $+\bm{k}$ mode be the subsystem $A$, and the $-\bm{k}$ mode the subsystem $B$. Then we have shown $\mathcal{S}[A,B]=0$ but $\mathcal{S}[A]=\mathcal{S}[B]>0$. The entropy of a subsystem is larger than the entropy of the combined system. These common properties are now often invoked in the discussions of entanglement entropy.


\section{Summary and Discussions}

We  conclude with a summary discussion of the three aspects of this work: objective, methodology and issues.

\subsection{Objective: Entropy of quantum fields and cosmological perturbations}

A central subject of interest in cosmology is structure formation. The classical theory  began  with Lifshitz's  gravitational perturbation theory in 1946.  Another important subject is  particle creation from the quantum vacuum, believed to be abundant at the Planck time,  first explored by Parker in 1966.  Both aspects enter in theories of cosmological perturbations from inflationary cosmology, in 1982,  where a quantum field is supposed to drive the universe into inflation and its fluctuations engender structures. 

The entropy budget of the universe generated from various  sources  is also an essential concern in cosmology. Main focus was placed on particle processes at various stages of the cosmic evolution. Entropy associated with cosmological perturbations met with a new level of challenge after inflation took center stage.  A necessary ingredient is to understand the entropy of quantum field processes, for free \cite{HuPav} and interacting fields \cite{HuKan}. That began in 1986.   Incorporating both the quantum theory of gravitational perturbations \cite{FMB}  and the ideas about the entropy of quantum fields into a theory for the entropy of cosmological perturbations took shape in 1993 in the work of \cite{BMP}.  

Our present work attempts to provide a synthesis of all the essential elements on this subject matter since that time,  constructed from a  more rigorous and comprehensive  theoretical framework based on the nonequilibrium quantum field theory of squeezed quantum systems.  We highlight its  advantages as follows.

\subsection{Methodology:  Nonequilibrium dynamics of squeezed quantum systems}

The technical core of this theory as applied to the present problem was presented in Sec. V,  VI and VIII while the results meeting our objective,  in Sec. VII.   The systematic theoretical development of the formalism can be found in 
\cite{CalHu08,PRE18,NEqFE,FDRSq}. Applying this theory for squeezed open quantum systems, as long as they are   Gaussian, enables us to obtain exact solutions for the fully nonequilibrium time-evolution of the cosmological perturbations, up to the Gaussian level.  

We mention a few specific advantages:    Essentially all quantities of  interest can be expressed in terms of two fundamental solutions of the equation of motion of the system plus initial conditions. This make it computationally efficient, and easier to identify the structures of the models, compared with some earlier approaches, say~\cite{InfSqV}. With an exact formulation one can leave the necessary mathematical approximations or physical procedures, such as  the regularization of  the stress tensor associated produced particles' density to the very end of the calculations where the actual expression is finally need.   

A very useful tool special to Gaussian systems is the covariance matrix, whose power lies in the fact that the calculations involving the infinite-dimensional density matrix elements can be reduced to those of a handful of covariance matrix elements. The authors of \cite{CamPar} have used them to compute the von Neumann entropy of the field, but they can be also be used to form the symplectic invariants~\cite{addesso} and the observables of the Gaussian systems. These physical quantities are very useful for  various applications in the context of open quantum systems like dynamical equilibration~\cite{CPR}, quantum entanglement~\cite{HHPRD}, nonequilibrium quantum thermodynamics~\cite{GTOS,PRE18}.  

In this work we follow the same spirit.  We first express the quantities of  interest, not just limited to the von Neumann entropy, in terms of the covariance matrix elements. We then will express these elements in terms of the aforementioned fundamental solutions. One can break down a  complex calculation to modular forms,  which makes the comparisons of results from different theories easier to identify, and be able to go further as needed, sailing along guided by the formalism . 

\subsection{Issues: Quantum correlations, coherence, phases}

Entropy generation from quantum cosmological perturbations, as  explained above, is weighted upon the  issue of entropy of quantum fields associated with vacuum particle creation.  Cosmological expansion having the same effect on a quantum state as squeezing, the issue becomes that of entropy generation from a squeezed quantum system. This is a good entry point to dissect the evolving explanations of  entropy generation from quantum cosmological perturbations. We also see this as a good intellectual exercise which can reveal the subtleties of this issue,  as it bears on the concepts of quantum correlations,  coherence and what (not) to do with the phase.

Intuitively one may easily attribute entropy generation to the particle number from pair creation,  as a result of the parametric amplification of the vacuum.   However, an actual calculation shows that,  starting in a vacuum state,  the field will end up in the two-mode squeezed vacuum stage after the parametric process, which is a pure state and no entropy change.  The authors of~\cite{HuPav} pointed out that if one chooses to describe the particle creation process in a Fock representation and totally ignores their phase relations one would come up with this conclusion.  Entropy is `generated' because one choose to ignore certain information of the field which could be just as important.  How one deals with the phase information is a different story.  A common argument   
is to appeal to the random phase approximation (RHA) which assumes that rapid phase oscillation tends to smear out its own footprints.  However the relevant phases are rather illusive. Let us examine this issue more closely.

Phase information can be placed in two categories: 1) the phase acquired during the evolution residing in  the changes of the rotation phase plus the squeeze phase read off from the Bogoliubov coefficients;   2) initial phase of the squeeze parameter or the state. The former is fixed and dynamically determined, but the latter is largely unknown and in principle cannot be canceled as such. Thus the information of the latter is mostly inaccessible. The authors of \cite{KME97}  concretized this by writing down the quantum Vlasov equations of the particle number density and the coherence of the created particles, in an attempt to establish the connection between quantum coherence and the  particle creation process which is non-Markovian. They argued that a nonzero entropy production associated with particle creation can emerge, if the variables associated with rapid phases are either plainly discarded or averaged out. This strategy is also adopted in~\cite{Prok93, BMP}.  However, this way to resolve the issue is not completely satisfactory in the sense that 1) both the particle number density and coherence may be oscillatory in time for the parametric process, so even though the coherence may oscillate more rapidly than the particle number does, the difference is not always dramatic enough to justify discarding the coherence;  2) The authors of~\cite{LCH10} show in an example that coherence does not always oscillate;  and 3) the low momentum modes have slow oscillation to make cancellation less effective.  In addition, resorting to phase cancellation over all modes does not apply here because the emergent entropy occurs in each mode. 

There is no denying that coherence tends to cancel because it is not positive definite as is the particle number density. Nonetheless it seems more reasonable to go after the coherence rather than  the phases literally. The authors of \cite{LCH10} further argued that behind the coherence between pair-produced particles, more precisely, there is quantum entanglement between them at play. They showed the clear presence of  entanglement by the von Neumann entropy of the reduced system after the degree of freedom of one party in the created particle pair is completely traced out. One can treat this as the emergent entropy associated with particle creation once one completely ignores or lose track of the information of one party in the particle pair. This we view as a better explanation of the root cause of emergent entropy associated with particle creation, and by extension, with  cosmological perturbations involving quantum fields like in the inflationary universe. 

We regained the result in~\cite{LCH10} using a totally different formulation, based on the nonequilibrium dynamics of squeezed quantum fields.  
A quantum description is necessary for addressing the foundational issues of quantum decoherence and quantum entanglement, and a nonequilibrium dynamics formulation is  necessary for time-dependent situations such as in a cosmological evolution.  In relation to this we showed that  a classical stochastic description of the quantum field used in~\cite{BMP}, though attractively simple, is inadequate to capture these important quantum attributes fundamental to quantum information issues. These authors construct a classical probability distribution function of the classical stochastic field, and the statistical nature of the field is studied through taking suitable statistical averages.  In contrast, we provided a fully quantum treatment by constructing the Wigner function of the quantum parametric field from  scratch.  We show that their classical  probability distribution function resembles our  Wigner function in form, but in contrast, the Wigner function, known to possess  the full information contained in a density matrix, is non-positive definite. That is where the quantum features of the system reside. \\

\noindent\textbf{Acknowledgment}  J.-T. Hsiang is supported by the Ministry of Science and Technology of Taiwan under Grant No.~MOST 110-2811-M-008-522.


\end{document}